# Statistical Power for Estimating Treatment Effects Using Difference-in-Differences and Comparative Interrupted Time Series Estimators with Variation in Treatment Timing

Peter Z. Schochet, Ph.D.
Senior Fellow
Mathematica
P.O. Box 2393
Princeton NJ, 08542-2393
pschochet@mathematica-mpr.com

October 2021

## Abstract

This article develops new closed-form variance expressions for power analyses for commonly used difference-in-differences (DID) and comparative interrupted time series (CITS) panel data estimators. The main contribution is to incorporate variation in treatment timing into the analysis. The power formulas also account for other key design features that arise in practice: autocorrelated errors, unequal measurement intervals, and clustering due to the unit of treatment assignment. We consider power formulas for both cross-sectional and longitudinal models and allow for covariates. An illustrative power analysis provides guidance on appropriate sample sizes. The key finding is that accounting for treatment timing increases required sample sizes. Further, DID estimators have considerably more power than standard CITS and ITS estimators. An available Shiny R dashboard performs the sample size calculations for the considered estimators.

JEL: C12, C13, C90

Keywords: Difference-in-differences designs, interrupted time series designs, statistical power, treatment effect estimation, variation in treatment timing

# 1. Introduction

When randomized controlled trials are not feasible and time series data are available, panel data methods can be used to estimate treatment effects on outcomes, by exploiting variation in policies and conditions over time and across locations. A feature of panel data is that treatment timing often varies across the sample, for example, due to differences across locations in treatment implementation (such as school turnaround efforts [Redding & Nguyen, 2020]), adoption of laws, or natural phenomena (such as the onset of Covid-19). Only recently has this timing feature been addressed in the literature for estimating treatment effects using difference-in-differences (DID) estimators but not for other panel data estimators (Athey & Imbens, 2018; Borusyak & Jaravel, 2017; Callaway & Sant'Anna, 2020; de Chaisemartin & D'Haultfœuille, 2020; Goodman-Bacon, 2018; Sun & Abraham, 2020). Another feature of panel data is that outcomes are often autocorrelated over time, which if not addressed can lead to estimated standard errors that are seriously biased downwards (Bertrand et al., 2004).

This article discusses new power formulas to assess required sample sizes that account for both of these real-world features for two classes of commonly used panel data estimators: (1) DID estimators that contrast changes in outcomes during the pre- and post-treatment periods across the treated and comparison groups, and (2) comparative interrupted time series (CITS) and associated ITS estimators that rely on fitted pre- and post-period trendlines. We adopt common specifications found in the literature for these estimators and build on the much smaller associated literature on statistical power that has focused only on settings without variation in treatment timing. We consider binary treatments—where individuals or groups who become treated remain so during the following periods—and focus on both continuous outcomes (such as student test scores) and binary outcomes aggregated as rates to the cluster level.

Another contribution of this work is that we incorporate recent approaches for adjusting for the clustering of individuals within groups—such as geographic or educational units—an issue where there has been considerable confusion in the causal inference literature for quasi-experimental designs (QEDs) (Abadie et al., 2017). We consider panel designs with separate cross-sections of individuals as well as longitudinal designs where the same individuals are followed over time. Further, unlike previous power analysis studies, we allow for time periods to be unevenly spaced and for the inclusion of model covariates to adjust for potential confounding bias and to improve precision. Using an event history approach, we develop power formulas for both point-in-time effects (to examine treatment dynamics) and pooled effects averaged over the full post-period. We use a potential outcomes framework (Holland, 1986; Rubin, 1974, 2005) to facilitate comparisons across the estimators. Our approach accommodates studies using either individual- or cluster-level (aggregate) data. A free Shiny R dashboard, *Power_Panel* (*add link when article is finalized*), performs the sample size calculations for the considered estimators.

The rest of the article is in eight sections. Section 2 discusses the related power analysis literature and Section 3 discusses our statistical power framework. In Section 4, we discuss clustering and Section 5 presents notation and target impact parameters. Sections 6 and 7 present sample size formulas for the DID and CITS/ITS estimators, where our main results are presented in theorems. In Section 8, we present an illustrative power analysis for the considered estimators, and then conclude. For reference, Table 1 lists the equation numbers for the variance formulas presented in the text for each estimator. Appendix A provides more details on the formulas.

## 2. Related Literature

We extend the previous power analysis literature for DID estimators to simultaneously allow for variation in treatment timing (not previously addressed), autocorrelated errors, clustering, both average and point-in-time effects, and uneven time intervals and covariates (both not previously addressed). Frison and Pocock (1992) provide sample size formulas for non-clustered, longitudinal DID designs with constant correlations across time, but do not consider clustered designs. McKenzie (2012) considers similar non-clustered designs with more general error structures, but do not provide explicit variance formulas. While Somers et al. (2013) consider clustered designs, they examine two time periods only and do not address staggered timing or autocorrelation. Burlig et al. (2019) provide sample size formulas for cross-sectional models with both clustering and autocorrelated errors for a more restrictive model specification than ours, but do not allow for staggered timing, uneven time intervals, covariates, or point-in-time effects.

To our knowledge, Bloom (1999) is the only article that has developed statistical power formulas for CITS and ITS estimators. However, this study only considers models with discrete post-period indicators, but not more common specifications that model both pre- and post-period trendlines that are considered here. Further, Bloom (1999) considers a single follow-up period only and does not consider staggered timing, autocorrelation, or covariates. Zhang et al. (2011), Kontopantelis (2018), Hawley et al. (2019), and Liu et al. (2020) use simulations to calculate power for the ITS design, but do not develop power formulas or address staggered timing. Several articles conduct within-study comparisons using RCT data to empirically compare RCT impact estimates with those obtained using CITS or ITS methods (Baicker & Svoronos, 2019; St. Clair et al., 2016), but do not develop variance formulas. In the theory sections, we show how our general formulas reduce to those from the literature.

### 3. Statistical Power Framework

Our power analysis relies on formulas for minimum detectable impacts, which represent the smallest true impacts that can be detected with a high probability. We scale the minimal detectable impacts into effect size (standard deviation) units—hereafter referred to as $MDE$s—which is common in policy research, especially for outcomes that are difficult to interpret in nominal units (although this scaling is not necessary). This approach parallels the statistical power literature for randomized controlled trials (RCTs) (see, e.g., Bloom, 1995; Donner & Klar, 2000; Murray, 1998; Schochet, 2008) and regression discontinuity designs (Schochet, 2009).

All the considered impact estimators are multiple regression estimators. Accordingly, relying on asymptotic normality and a classical hypothesis testing approach, we assume the use of $t$-statistics to test null hypotheses of the form, $H_0: \beta = 0$, where $\beta$ is the impact parameter of interest. This approach yields the following $MDE$ formula for a two-tailed hypothesis test:

$$MDE = \left(\frac{1}{\sigma}\right) Factor(\alpha, \lambda, df) \sqrt{Var(\hat{\beta})}, \tag{1}$$

where $Factor(\alpha, \lambda, df) = [T^{-1}(\alpha/2, df) + T^{-1}(\lambda, df)]$; $T^{-1}$ is the inverse of the student's $t$ distribution function; $\alpha$ is the significance level; $\lambda$ is the statistical power level; $df$ is the degrees of freedom; $\hat{\beta}$ is the impact estimator; $Var(\hat{\beta})$ is the variance of $\hat{\beta}$ based on the specific design features; and $\sigma$ is the standard deviation of the outcome variable conditional on the model covariates. In our application, $\sigma$ represents cluster-level variation at a particular time point.

Alternatively, one can solve (1) to calculate required sample sizes or the number of time periods to achieve a given $MDE$, assuming specific values of $\alpha$, $\lambda$, and the design parameters in $Var(\hat{\beta})$. Study sample sizes enter (1) primarily through $Var(\hat{\beta})$ but also through $Factor(\alpha, \lambda, df)$ because of $df$. Hill et al. (2008) and Lipsey et al. (2012) discuss a framework for selecting $MDE$ targets for education evaluations, for example, by examining the natural growth in outcomes over time, policy-relevant gaps across subgroups, and observed effect sizes from previous similar evaluations.

The power formula in (1) hinges critically on $Var(\hat{\beta})$. Thus, this article focuses on calculating closed-formed expressions for $Var(\hat{\beta})$ for the considered designs. However, we begin by discussing the general issue of clustering that can have a large effect on precision.

## 4. Clustering

Clustering occurs when the outcomes of individuals (or broader units) in the dataset are correlated. Consider a panel analysis using data on the same individuals (or units) over time. In this case, outcomes are likely to be correlated for the same individual (or unit) across time periods, and our variance formulas incorporate this form of clustering. However, there has been much confusion in the QED causal inference literature about how to account for clustering across different individuals (or units) in the sample (Abadie et al., 2017).

In this article, we adopt an approach developed for RCTs and inspired by design-based theory to account for clustering across sample members or units (Abadie et al., 2017; Freedman, 2008; Imbens & Rubin, 2015; Schochet, 2010, 2013, 2016; Schochet et al., 2020). Under this approach, one must be explicit about what is random under repeated sampling. For instance, if the study units are assumed to be fixed (which occurs in finite population settings), the main source of randomness is the treatment assignment itself. Accordingly, in this setting, the presence of clustering hinges on the considered unit of treatment assignment.

To determine this unit, it is useful to consider what would be the parallel unit under a hypothetical RCT. To help fix concepts, consider a panel data study examining whether Covid-19 had a larger effect on students in states where the pandemic hit earlier than in states where it hit later. In this case, the hypothetical unit of treatment assignment under an RCT would be the state, and thus, similarly in the panel data context. However, suppose the study instead examined the effects of pandemic-related educational policies (such as distance learning) that differed by school district. In this case, we would treat districts as the unit of treatment assignment and the source of clustering.

Note that in some designs, study units are formally sampled for the study from broader populations (e.g., studies using national datasets with multi-stage sample designs or multi-site random block designs) or are deemed to have been so. In these cases, study results are assumed to generalize beyond the study sample, and clustering from the sampling of units becomes pertinent (Schochet, 2008). We do not formally consider such forms of clustering because their variance estimators differ from those developed here, which is due to the emergence of correlations between the outcomes of the treatment and comparison groups within the same higher-level clusters. However, our variance formulas are conservative in these cases (because they ignore these correlations). Further, our approach applies to settings where the higher-level clusters are treated as fixed effects rather than as random blocks.

Clustering increases the $MDE$s in (1) for two interrelated reasons (Murray, 1998; Donner & Klar, 2000; Hedges, 2007; Schochet, 2008). First, design effects due to correlated outcomes reduce effective sample sizes, and hence precision, as can be quantified by intraclass correlation coefficients ($ICC$s). Second, $df$s are based on the number of clusters, not the number of individuals, which increases $Factor(\alpha, \lambda, df)$ values in (1).

While many designs in our setting are likely to be clustered, there may be longitudinal designs where the unit of treatment assignment is the individual. For instance, consider a study comparing the effects of Covid-19 on students in a school district with and without internet access for distance learning. In this case, the unit of treatment assignment could be the student. While we consider more general clustered designs, our variance formulas also apply (reduce) to non-clustered, longitudinal designs by setting the pertinent $ICC$ parameters to 0 and treating clusters as individuals (as discussed further in Section 6).

## 5. Definitions, Assumptions, and Target Impact Parameters

We consider a clustered panel data setting with $M$ total clusters defined by the unit of treatment assignment. We assume $M_T$ clusters in the treatment group and $M_C$ untreated clusters in the comparison group. We assume the same clusters remain in the sample for each of $P$ time periods (e.g., where time is measured in months, semesters, or years). We denote time periods by $t = 1, \ldots, P$ with both pre- and post-treatment periods. As discussed later, the time periods do not necessarily have to be evenly spaced, but we avoid this notational inconvenience for now.

We assume $K$ treatment timing groups in total, with treatment start times of $S_k$ for timing group $k \in \{1, \ldots, K\}$ ordered from earliest to most recent, with $M_{Tk}$ treatment clusters in group $k$. We assume $M_{Ck}$ comparison clusters are matched to timing group $k$ ($M_{Tk}$ and $M_{Ck}$ can differ). We do not consider an analysis where each timing group is compared to the full comparison group (as is sometimes done in the literature), because under our event history framework this approach considerably complicates the variance formulas due to covariance terms that arise across the timing group estimators. It can be shown, however, that the power formulas developed here are good approximations for panel designs with an overlapping comparison group sample (e.g., by proportionally splitting it across the timing groups). Our stratified approach may also be desirable in practice, especially if there are differences in cluster-level characteristics across the timing groups (see Haw & Hatfield, 2020 for a discussion of matching issues in panel studies).

We assume multiple time periods and the availability of both pre- and post-period data ($S_k \geq 2$ for the DID design and $4 \leq S_k \leq (P - 2)$ for the CITS and ITS designs for all $k$. We assume treatment effects can be observed and measured *starting* in period $S_k$, which pertains to the *first* post-treatment period. Let $T_j$ be the treatment indicator that equals 1 for ever-treated clusters and 0 for comparison (never-treated) clusters. Further, let $G_{Tj} = k$ denote the group of treatment clusters in timing group $k$ (with onset time $S_k$), where $G_{Cj} = k$ are the group of matched comparison clusters. Finally, let $A_k = (P - S_k + 1)$ denote the number of post-intervention ("after") periods, where $B_k = (S_k - 1)$ are the number of pre-intervention ("before") periods. For reference, Table 2 provides key parameter definitions used throughout the article.

We assume data are available for individuals in each study cluster at each time period, which allows us to fully examine sources of variation in the model error terms. However, the power formulas apply also to analyses using data averaged to the cluster level or to lower-level units within the study clusters. Data aggregation (averaging) aligns with our framework because we consider *linear* panel data estimators, so the regression model structure is not affected by aggregation (Greene, 2012; Schochet, 2020). The analysis, however, cannot be conducted if data are available only for units at higher levels than the unit of treatment assignment.

We index individuals by $ijt$ for individual $i$ in cluster $j$ in time $t$. For each cluster, either all individuals are treated or not. For simplicity and as customary in the power literature, we assume a balanced design with $N_{jt} = N$ individuals per cluster in each time period; for unbalanced designs, the average sample size across all time periods and clusters, $\bar{\bar{N}} = \frac{1}{PM} \sum_{t=1}^{P} \sum_{j=1}^{M} N_{jt}$, can be used in the power formulas as an approximation (Donner & Klar, 2000; Kish, 1995; Murray, 1998; Schochet, 2008). This assumption does not materially affect the analysis as power for clustered designs is determined primarily by the number of clusters, not the number of individuals per cluster (unless this number is small). We allow for both separate cross-sections of individuals (e.g., separate cohorts of fourth graders in each time period) and longitudinal data where the same persons are followed over time.

To make the large amount of notation concrete, consider the panel study conducted by Zimmer et al. (2017) that estimated the effects of school turnaround strategies in low-performing Tennessee schools on student test scores. In this study, there were $P = 8$ time periods with $M_T = 51$ schools (clusters) in the treatment group and $M_C = 28$ comparison clusters. There was variation in treatment timing, with 19 treatment schools treated at time point $t = 6$, 20 schools treated at $t = 7$, and 12 schools treated at $t = 8$. Thus, the study had three timing groups ($k = 1$, 2, and 3) with associated treatment start times of $S_1 = 6$, $S_2 = 7$, and $S_3 = 8$ and treatment cluster sample sizes of $M_{T1} = 19$ in timing group 1 ($G_{Tj} = 1$), $M_{T2} = 20$ in timing group 2 ($G_{Tj} = 2$), and $M_{T3} = 12$ in timing group 3 ($G_{Tj} = 3$). The lengths of the pre- and post-periods were $B_1 = 5$ and $A_1 = 3$ for $k = 1$; $B_2 = 6$ and $A_2 = 2$ for $k = 2$; and $B_3 = 7$ and $A_3 = 1$ for $k = 3$. Finally, there was an average of $N = 230$ students per school per time period. We use this running example throughout the article.

We invoke several assumptions that apply to both the DID and CITS/ITS estimators to obtain unbiased estimators of well-defined treatment effect parameters, where additional assumptions specific to each design are presented in Sections 6 and 7:

**Assumption 1. The stable unit treatment value assumption (SUTVA) (Rubin 1986).** Let $Y_{ij}(\mathbf{T}_{clus})$ denote the potential outcome for an individual given the random vector of all cluster treatment assignments, $\mathbf{T}_{clus}$. Then, if $T_j = T_j'$ for cluster $j$, we have that $Y_{ij}(\mathbf{T}_{clus}) = Y_{ij}(\mathbf{T}_{clus}')$.

SUTVA allows us to express $Y_{ij}(\mathbf{T}_{clus})$ as $Y_{ij}(T_j)$, so that the potential outcomes of an individual in cluster $j$ depends only on the cluster's treatment assignment and not on the treatment assignments of other clusters in the sample. More specifically, SUTVA allows us to define $Y_{ijt}(1)$ as the potential outcome for the individual in the treatment condition and $Y_{ijt}(0)$ as the potential outcome in the non-treated condition. Potential outcomes are assumed to be continuous variables, although our methods also approximately apply to binary outcomes that are aggregated as rates or proportions to the cluster level. Using this framework, the data generating process for the *observed* outcome at each time point, $y_{ijt}$, can be expressed as $y_{ijt} = T_j Y_{ijt}(1) + (1 - T_j) Y_{ijt}(0)$. SUTVA also requires that units cannot receive different forms of the treatment.

**Assumption 2. No anticipatory behavior.** We assume that a future treatment does not affect past outcomes, so that $y_{ijt} = Y_{ijt}(1) = Y_{ijt}(0)$ for $t < S_j$. This assumption is required to rule out situations where individuals or units change their behavior in anticipation of a treatment that is likely to occur in the future, which could influence pre-period outcomes.

To define the focal treatment effect parameters for our analysis, we adopt an "event history" approach that aggregates treatment effect parameters across particular treatment timing groups and time periods (Callaway and Sant'Anna, 2020; Sun and Abraham, 2020). With variation in treatment timing, an important consideration for defining our focal impact parameters is whether to measure the post-program impacts in calendar time or relative to exposure to the treatment. Consider first a calendar-time analysis, where we define $ATT_{kq}$ as the average effect of treatment on the treated ($ATT$) for clusters in timing group $G_{Tj} = k$ in post-period $q$ (assuming $q \geq S_k$):

$$ATT_{kq} = E\big(\bar{Y}(1)_{\{t=q\}} - \bar{Y}(0)_{\{t=q\}}\big| G_{Tj} = k\big). \qquad (2)$$

Here, $E(\bar{Y}(1)_{\{t=q\}}|G_{Tj} = k) = \frac{1}{M_{Tk}N} \sum_{j:G_{Tj}=k}^{M_{Tk}} \sum_{i=1}^{N} E(Y_{ijq}(1))$ is the mean cluster-level outcome in the treated condition at time $q$ for timing group $k$, and similarly for $E(\bar{Y}(0)_{\{t=q\}}|G_{Tj} = k)$ in the untreated condition. Following Callaway and Sant'Anna (2020), our first impact parameter of interest averages the $ATT_{kq}$ parameters across timing groups:

$$ATT_q = \frac{1}{\sum_{k=1}^{K} w_{kq} I(q \geq S_k)} \sum_{k=1}^{K} w_{kq} I(q \geq S_k) ATT_{kq}, \qquad (3)$$

where $I(q \geq S_k)$ is an indicator that equals 1 if $q$ is a post-period for timing group $k$ and 0 otherwise, and $w_{kq}$ are weights (which we set to 1, as discussed in Section 6.1). Some timing groups may not contribute to (3). For instance, in our running example, $q = 7$ is a post-period for timing groups $k = 1$ and $k = 2$ with $S_1 = 6$ and $S_2 = 7$, but not for timing group $k = 3$ with $S_3 = 8$, so timing group 3 would be excluded from (3) to calculate $ATT_7$.

Note that at calendar time $q$, exposure to the intervention, $(q - S_k + 1)$, will *differ* across timing groups, which could complicate the interpretation of the $ATT_q$ parameters if impacts vary over time. Using our running example, at $q = 8$, the exposure time is 3 periods for timing group 1, compared to only 2 periods for timing group 2 and 1 period for timing group 3. Thus, an alternative approach used by Sun and Abraham (2020) is to realign the data to measure post-period impacts relative to treatment exposure, where exposure time, $t_e$, can be mapped to calendar time using $t_e = (t - S_k + 1) = (t - B_k)$. This approach yields the following $ATT_l^e$ parameter pertaining to impacts at exposure point $l$ that are averaged over the timing groups:

$$ATT_l^e = \frac{1}{\sum_{k=1}^{K} w_{kl}^e I(l \leq A_k)} \sum_{k=1}^{K} w_{kl}^e I(l \leq A_k) ATT_{kl}^e, \qquad (4)$$

where $ATT_{kl}^e = E\big(\bar{Y}(1)_{\{t_e=l\}} - \bar{Y}(0)_{\{t_e=l\}} \big| G_{Tj} = k\big)$ is the $ATT$ parameter for timing group $k$ at exposure point $l$; $I(l \leq A_k)$ is an indicator that equals 1 if the length of the post-period ($A_k$) is at least $l$ periods and 0 otherwise; and $w_{kl}^e$ are weights (which we set to 1). These point-in-time $ATT_l^e$ parameters are useful for examining treatment dynamics (Sun and Abraham, 2020), although potential differences in impacts across timing groups could affect interpretation.

Power concerns are similar for the point-in-time $ATT_q$ and $ATT_l^e$ parameters as they both rely on the same $ATT_{kq}$ components but are organized and averaged in different ways. Thus, our analysis applies to both parameters. While we prefer the $ATT_l^e$ parameter (since it has a clearer interpretation), for ease of exposition, we use calendar time notation because it conforms to the format of time series data that are typically available for analysis; we then transform the results into exposure time as needed. The choice of notation does not change the results.

Another key impact parameter for our analysis, $ATT_{POOL}$, takes a weighted average of either the $ATT_q$ or $ATT_l^e$ parameters over their respective post-periods to obtain pooled treatment effects averaged over the full observed post-period (both approaches yield the same result). This pooled parameter has policy relevance as a summary measure of treatment effects, for instance, for use in benefit-cost analyses. An intuitive way to express the $ATT_{POOL}$ parameter is as follows:

$$ATT_{POOL} = \frac{1}{\sum_k \sum_q w_{kq}} \sum_{k=1}^{K} \sum_{q=S_k}^{P} w_{kq} ATT_{kq}, \tag{5}$$

where we set $w_{kq} = 1$ for our analysis.[1] This aggregate estimand is flexible in that it allows for heterogenous treatment effects both across timing groups and across time within timing groups.

The estimation challenge for the considered treatment effect parameters is that we do not observe counterfactual outcomes for the treated groups during the post-treatment period, that is, the $E(\bar{Y}(0)_{\{t=q\}}|G_{Tj} = k)$ terms in (2). Thus, our considered DID and CITS/ITS panel methods estimate these counterfactuals under various identification assumptions, which then allows for consistent estimation of the $ATT_q$, $ATT_l^e$, and $ATT_{POOL}$ parameters in (3), (4), and (5).

## 6. Difference-in-Differences (DID) Estimator

DID methods identify causal effects from panel data by contrasting changes in outcomes during the pre- and post-treatment periods across the treatment and comparison groups. The large literature on DID methods focuses on designs without variation in treatment timing (e.g.,

---

[1]Equivalently, we can define $ATT_{POOL}$ using $ATT_{POOL} = \frac{1}{\sum_q w_q} \sum_{q=min_k(S_k)}^{P} w_q\, ATT_q = \frac{1}{\sum_l w_l^e} \sum_{l=1}^{max_k(A_k)} w_l^e ATT_l^e$, where $w_q = \sum_{k=1}^{K} w_{kq} I(q \geq S_k)$ and $w_l^e = \sum_{k=1}^{K} w_{kl}^e I(l \leq A_k)$ are weights.

Ashenfelter, 1978; Ashenfelter & Card, 1985; Bertrand et al., 2004; Angrist & Pischke, 2009). However, a smaller recent literature (cited in the introduction) considers impact estimation with staggered treatment timing to produce unbiased estimators. This literature underlies our analysis.

The key identifying assumption for DID methods is that in the absence of treatment, the mean outcomes for the treatment and comparison groups would exhibit parallel trends over time (Heckman et al., 1997; Abadie, 2005):

**Assumption DID.1. Parallel trends.** For each timing group, $k \geq 1$, mean counterfactual outcomes for the treatment and comparison groups exhibit parallel trends for each post-period, $q \geq S_k$, relative to the average pre-period:

$$E\big(\bar{Y}(0)_{\{t=q\}} - \bar{\bar{Y}}(0)_{\{t<S_k\}}\big|G_{Tj} = k\big) = E\big(\bar{Y}(0)_{\{t=q\}} - \bar{\bar{Y}}(0)_{\{t<S_k\}}\big|G_{Cj} = k\big). \qquad (6)$$

Here, $E(\bar{\bar{Y}}(0)_{\{t<S_k\}}|G_{Tj} = k) = \frac{1}{B_k}\sum_{b=1}^{B_k} E(\bar{Y}(0)_{\{t=b\}}|G_{Tj} = k)$ is the mean counterfactual outcome averaged over all pre-periods and all treatment clusters in timing group $k$ (using equal pre-period weights), and similarly for $E(\bar{\bar{Y}}(0)_{\{t<S_k\}}|G_{Cj} = k)$ for the comparison clusters.

Intuitively, this assumption allows us to obtain an unbiased estimate of the unobserved counterfactual, $E\big(\bar{Y}(0)_{\{t=q\}}\big|G_{Tj} = k\big)$, using $(\bar{\bar{y}}_{\{t<S_k\}}^{G_{Tj}=k} + \bar{y}_{\{t=q\}}^{G_{Cj}=k} - \bar{\bar{y}}_{\{t<S_k\}}^{G_{Cj}=k})$, where $\bar{\bar{y}}_{\{t<S_k\}}^{G_{Tj}=k} = \frac{1}{B_k M_{Tk} N}\sum_{t=1}^{B_k}\sum_{j:G_{Tj}=k}^{M_{Tk}}\sum_{i=1}^{N} y_{ijt}$ is the observed pre-period mean outcome for treatments in timing group $k$, and similarly for $\bar{y}_{\{t=q\}}^{G_{Cj}=k}$ and $\bar{\bar{y}}_{\{t<S_k\}}^{G_{Cj}=k}$; this yields the DID estimator discussed below.

In what follows, we develop variance formulas for the DID estimator, first for the cross-sectional design and then for the longitudinal design.

*6.1. Cross-Sectional Analysis: Framework*

To consider DID impact and associated variance estimators for the $ATT_q$, $ATT_l^e$, and $ATT_{POOL}$ parameters, we rely on the following regression model, adapted from the literature to our context, using stacked data on separate cross-sections of individuals nested within study units (clusters) over time:

$$y_{ijt} = \alpha_j + \sum_{k=1}^{K} \delta_{kt} I(G_j = k) + \sum_{k=1}^{K} \sum_{q=S_k}^{P} \beta_{kq} I(G_j = k) F_{ijt,q} T_j + \theta_{jt} + \varepsilon_{ijt}; \tag{7}$$

$$\theta_{jt} = \rho \theta_{j(t-1)} + \eta_{jt}.$$

In this model, $\alpha_j$ and $\delta_{kt}$ are cluster and time fixed effects, $I(G_j = k)$ is an indicator that equals 1 if either $G_{Tj} = k$ or $G_{Cj} = k$ and 0 otherwise, and $F_{ijt,q}$ is a period $q$ indicator that equals 1 if $t = q$ and 0 otherwise. The random error terms are assumed to have mean zero and to be distributed independently of each other: $\theta_{jt}$ captures the correlations of individuals within the same cluster and time period, and $\varepsilon_{ijt}$ are $iid$ $(0, \sigma_\varepsilon^2)$ individual-level errors. Following the influential work of Bertrand et al. (2004), we allow $\theta_{jt}$ to be correlated over time using an autoregressive process of order 1 (AR(1)), where $|\rho| < 1$ is the autocorrelation parameter and $\eta_{jt}$ are $iid$ $(0, \sigma_\eta^2)$ errors with $E(\eta_{jt} \theta_{j(t-1)}) = 0$. For our analysis, we assume the same autocorrelation structure during the pre- and post-treatment periods and for each timing group and their associated comparison group. Under the AR(1) model with long panels, $\sigma_\theta^2 = Var(\theta_{jt}) = \sigma_\eta^2/(1 - \rho^2)$, which we assume hereafter.[2] This is not a restrictive assumption as $Var(\theta_{jt})$ converges quickly over time (e.g., for $\rho = 0.5$, the variance stabilizes after 3 periods).

Under the AR(1) model, correlations are larger for cluster observations closer in time than further apart, so $Corr(\theta_{jt}, \theta_{j(t-p)}) = \rho^p$ for $p = 1, 2, \dots, (t-1)$. Accounting for autocorrelated errors in many settings is important because cluster-level outcomes are typically highly correlated over time, which if ignored can lead to serious underestimates of standard errors (Bertrand et al., 2004). For instance, pertinent to studies in education, when estimating (7) using National Assessment of Education Progress (NAEP) public use data on 28 school districts over 9 nine years (excluding the interaction terms), we find a $\rho$ value of 0.49 for 4th grade reading scores and 0.46 for 4th grade math scores (both are statistically significant at the 1 percent level). Thus, the autocorrelations are substantial, even after controlling for time fixed effects.

---

[2] This variance can be obtained by first noting that through back substitution, $\theta_{jt} = \rho^p \theta_{j(t-p)} + \sum_{k=0}^{p-1} \rho^k \eta_{j(t-k)}$. If we let $p \to \infty$, then $Var(\theta_{jt}) = \sigma_\eta^2 \sum_{k=0}^{\infty} \rho^{2k} = \sigma_\eta^2/(1 - \rho^2)$.

In DID settings, time intervals are often equally spaced, but not always. Thus, we consider general settings where time measurements can be unevenly spaced within timing groups but not across them. We follow Baltagi and Wu (1999) who discuss OLS estimation with unequal time intervals that fully maintains the AR(1) structure.[3] Under this approach, correlations between successive observations are based on their time differences (e.g., the correlation is $\rho^3$ for successive observations 3 time units apart). To use this approach, for each $t$, we use the notation, $Time_t$, to denote the elapsed time units between outcome measurement and a common reference point; thus, $t$ refers to a time label (counter), whereas $Time_t$ refers to elapsed calendar time.

The regression model in (7) includes three-way interactions between indicators of timing group, time period, and treatment status; the model, however, excludes interactions for each comparison group and pre-period. Thus, the resulting OLS estimators, $\hat{\beta}_{DID,kq}$, provide DID estimates for all post-treatment time points relative to the mean pre-period outcome. Formally, the DID estimator for timing group $k$ in time period $q \geq S_k$ is $\hat{\beta}_{DID,kq} = \left(\bar{y}_{\{t=q\}}^{G_{Tj}=k} - \bar{\bar{y}}_{\{t<S_k\}}^{G_{Tj}=k}\right) - \left(\bar{y}_{\{t=q\}}^{G_{Cj}=k} - \bar{\bar{y}}_{\{t<S_k\}}^{G_{Cj}=k}\right)$, where $\bar{y}$ and $\bar{\bar{y}}$ are mean observed outcomes defined above. For instance, using our running example, the DID estimator at time $q = 7$ for timing group $k = 1$ with $S_1 = 6$ can be calculated using $\hat{\beta}_{DID,17} = \left(\bar{y}_{\{t=7\}}^{G_{Tj}=1} - \bar{\bar{y}}_{\{t<6\}}^{G_{Tj}=1}\right) - \left(\bar{y}_{\{t=7\}}^{G_{Cj}=1} - \bar{\bar{y}}_{\{t<6\}}^{G_{Cj}=1}\right)$. However, the DID estimator, $\hat{\beta}_{DID,37}$, would not be germane for timing group $k = 3$ with $S_3 = 8$, because $q = 7$ is not a post-period for these clusters. Note that the OLS estimates using (7) are the same as those from models run separately by timing group.[4]

---

[3] This approach applies a data transformation using quasi-differences that yields independent observations, which generalizes the usual Prais–Winsten transformation, $y_{ijt} - \rho y_{ij(t-1)}$, for the model with even time spacing.

We can now aggregate the $\hat{\beta}_{DID,kq}$ estimators across timing groups to obtain an unbiased estimator for the calendar-time $ATT_q$ parameter in (3) using

$$\hat{\beta}_{DID,q} = \frac{1}{\sum_{k=1}^{K} w_{kq} I(q \geq S_k)} \sum_{k=1}^{K} w_{kq} I(q \geq S_k) \hat{\beta}_{DID,kq}. \quad (8)$$

Similarly, we can estimate the exposure-specific $ATT_l^e$ parameter in (4) using

$$\hat{\beta}_{DID,l}^e = \frac{1}{\sum_{k=1}^{K} w_{kl}^e I(l \leq A_k)} \sum_{k=1}^{K} w_{kl}^e I(l \leq A_k) \hat{\beta}_{DID,kq^e}, \quad (9)$$

where $q^e = (l + B_k)$ converts exposure time to calendar time to select the pertinent $\hat{\beta}_{DID,kq^e}$ estimators. Finally, we can aggregate $\hat{\beta}_{DID,q}$ or $\hat{\beta}_{DID,l}^e$ across their post-periods to obtain an unbiased estimator for the $ATT_{POOL}$ parameter in (5) using

$$\hat{\beta}_{DID,POOL} = \frac{1}{\sum_k \sum_q w_{kq}} \sum_{k=1}^{K} \sum_{q=S_k}^{P} w_{kq} \hat{\beta}_{DID,kq}, \quad (10)$$

or using the equivalent expressions in footnote 1 in Section 5.

For weighting, we follow Sun and Abraham (2020) who use $w_{kq} = w_{kl}^e = 1$, so that each included timing group receives the same weight in calculating the point-in-time estimators, $\hat{\beta}_{DID,kq}$. This weighting scheme also implies that when calculating $\hat{\beta}_{DID,POOL}$, the weight for each timing group is $A_k$, the length of its post-period (that is, , $w_k = \sum_{q=S_k}^{P} w_{kq} = A_k$). For instance, in our running example, timing group 1 with $A_1 = 3$ would receive a larger weight than timing group 2 with $A_2 = 2$ and timing group 3 with $A_3 = 1$, because it has more exposure periods. In this setting, $\hat{\beta}_{DID,POOL}$ can be interpreted as the average treatment effect observed in the sample. Other options exist for $w_{kq}$ and $w_{kl}^e$, such as weighting according to the number of clusters ($M_{Tk}$ and $M_{Ck}$) (Callaway and Sant'Anna, 2020) or inversely proportional to timing group variances.

---

[4] For estimation, to avoid perfect collinearity among the regressors in (7), it is necessary to omit some fixed effect parameters. A simple strategy is to omit the $\delta_{k1}$ parameter for each timing group as well as the intercept.

Finally, we note that our event history approach differs from the approach of Goodman-Bacon (2018) who examines impact estimation for a variant of model (7) where the three-way interaction terms are replaced by $\beta D_{jt}$, where $D_{jt}$ is an indicator that equals 1 if cluster $j$ in the treatment group was treated at or before time $t$ and 0 otherwise, and where a common comparison group is used across timing groups. Under this specification, $\beta$ is the average $ATT$ parameter over the post-period. With staggered treatment timing, Goodman-Bacon (2018) shows that the OLS estimator, $\hat{\beta}$, becomes a weighted average of separate two-by-two DID estimators, where each treatment timing group is compared not only to the common comparison group, but also to each other based on treatment timing. More specifically, a later timing group serves as a comparison group for the early timing group before its treatment begins and the early group then serves as a comparison group for the later timing group after its treatment begins.

We do not adopt this specification, however, because it only recovers the $ATT$ parameter when treatment effects are homogeneous across timing groups and over time (Callaway & Sant'Anna, 2020; de Chaisemartin & D'Haultfœuille, 2020; Sun & Abraham, 2020). These assumptions, however, are unrealistic in practice. Instead, our event history approach allows for both sources of heterogeneity and yields unbiased and more interpretable impact estimates.

*6.2. Cross-sectional Analysis: Variance Estimation*

In this section, we first obtain the variance formula for the pooled DID estimator, $\hat{\beta}_{DID,POOL}$ in (10) and then discuss how this formula reduces to the variance formulas for the point-in-time estimators, $\hat{\beta}_{DID,q}$ in (8) and $\hat{\beta}^e_{DID,l}$ in (9).

To obtain the variance of $\hat{\beta}_{DID,POOL}$, it is useful to express $\hat{\beta}_{DID,POOL}$ in (10) as an average of treatment effects across each timing group, $\hat{\beta}_{DID,k}$, as follows:

$$\hat{\beta}_{DID,POOL} = \frac{1}{\sum_{k=1}^{K} A_k} \sum_{k=1}^{K} A_k \hat{\beta}_{DID,k}, \tag{11}$$

where $\hat{\beta}_{DID,k} = \frac{1}{A_k} \sum_{q=S_k}^{P} \hat{\beta}_{DID,kq}$, recalling that our weighting scheme uses $w_{kq} = 1$ and $w_k = \sum_{q=S_k}^{P} w_{kq} = A_k$. Because of the assumed independence of the outcomes across the timing groups, the variance of $\hat{\beta}_{DID,POOL}$ can then be obtained using the following simple relation:

$$Var(\hat{\beta}_{DID,POOL}) = \frac{1}{(\sum_{k=1}^{K} A_k)^2} \sum_{k=1}^{K} A_k^2 Var(\hat{\beta}_{DID,k}). \tag{12}$$

We can now calculate $Var(\hat{\beta}_{DID,POOL})$ by first calculating $Var(\hat{\beta}_{DID,k})$ using the variances and covariances across the pre- and post-period means based on the model error structure in (7).

To fix concepts, we first consider the variance formula for timing group $k$ without the AR(1) structure, which can be expressed as follows:

$$Var(\hat{\beta}_{DID,k}) = \left(\frac{1}{M_{Tk}} + \frac{1}{M_{Ck}}\right)\left(\frac{1}{A_k} + \frac{1}{B_k}\right)\left(\sigma_\theta^2 + \frac{\sigma_\varepsilon^2}{N}\right). \tag{13}$$

This variance is intuitive in that it gets smaller as the number of clusters ($M_{Tk}$ and $M_{Ck}$), the number of individuals per cluster ($N$), and the lengths of the pre- and post-intervention periods ($B_k$ and $A_k$) increase. Clustering effects arise due to $\sigma_\theta^2$. If $\sigma_\theta^2 = 0$, (13) reduces to the non-clustered design.

If we now add the AR(1) structure, the calculations become considerably more complex, because the mean outcomes for a particular cluster become correlated across time periods (both within the pre- and post-periods as well as across them), where the correlations diminish over time. This leads to our first main variance result provided in the following new theorem.

**Theorem 1.** The variance of the pooled DID estimator, $\hat{\beta}_{DID,POOL}$, in (12) obtained from the model in (7) that incorporates clustering, variation in intervention timing, the AR(1) error structure, and general measurement intervals is as follows:

$$Var(\hat{\beta}_{DID,POOL})$$
$$= \frac{1}{(\sum_{k=1}^{K} A_k)^2} \sum_{k=1}^{K} A_k^2 \left(\frac{1}{M_{Tk}} + \frac{1}{M_{Ck}}\right) \left[\sigma_\theta^2 \left(\frac{1}{A_k} + \frac{1}{B_k} + \frac{(A_k - 1)}{A_k} \bar{\rho}_k^{Post}\right.\right. \quad (14)$$
$$\left.\left. + \frac{(B_k - 1)}{B_k} \bar{\rho}_k^{Pre} - 2\bar{\rho}_k^{Pre-Post}\right) + \frac{\sigma_\varepsilon^2}{N}\left(\frac{1}{A_k} + \frac{1}{B_k}\right)\right].$$

Here, $\bar{\rho}_k^{Post} = \frac{2}{A_k(A_k-1)} \sum_{a=S_k}^{P} \sum_{a'>a}^{P} \rho^{(Time_{a'} - Time_a)}$ denotes the average autocorrelation in the post-period; $\bar{\rho}_k^{Pre} = \frac{2}{B_k(B_k-1)} \sum_{b=1}^{B_k} \sum_{b'>b}^{B_k} \rho^{(Time_{b'} - Time_b)}$ denotes the average autocorrelation in the pre-period; $\bar{\rho}_k^{Pre-Post} = \frac{1}{B_k A_k} \sum_{b=1}^{B_k} \sum_{a=S_k}^{P} \rho^{(Time_a - Time_b)}$ denotes the average autocorrelation between the pre- and post-periods; and other terms are defined above.

In this expression, if $\rho > 0$ (the typical case), we find that the $\bar{\rho}_k^{Post}$ and $\bar{\rho}_k^{Pre}$ terms increase variance whereas the $\bar{\rho}_k^{Pre-Post}$ term reduces variance. In most cases, the $\bar{\rho}_k^{Post}$ and $\bar{\rho}_k^{Pre}$ terms will be larger than the $\bar{\rho}_k^{Pre-Post}$ term as cluster observations are, on average, closer in time within a pre- or post-period than across them. Thus, accounting for the AR(1) structure tends to reduce precision.

However, there are several cases where accounting for the autocorrelated errors improves precision. First, this will occur if we assume constant autocorrelations across time as is sometimes specified in the power literature for shorter panels (Frison & Pocock, 1992; McKenzie, 2012). In this case, $Corr(\theta_{jt}, \theta_{j(t-p)}) = \rho$, and the $Var(\hat{\beta}_{DID,k})$ formula reduces to $\left(\frac{1}{M_{Tk}} + \frac{1}{M_{Ck}}\right)\left(\frac{1}{A_k} + \frac{1}{B_k}\right)[\sigma_\theta^2(1-\rho) + \frac{\sigma_\varepsilon^2}{N}]$. Second, in short panels with $A_k = 1$ and $B_k = 1$, the $Var(\hat{\beta}_{DID,k})$ formula reduces to $\left(\frac{1}{M_{Tk}} + \frac{1}{M_{Ck}}\right)\left[2\sigma_\theta^2(1 - \bar{\rho}_k^{Pre-Post}) + \frac{2\sigma_\varepsilon^2}{N}\right]$, so the $\bar{\rho}_k^{Pre-Post}$ term improves precision in this case (a similar result holds if $A_k = 1$ or $B_k = 1$). Finally, if $\rho$ is close to 1, the $\bar{\rho}_k^{Pre-Post}$ term can offset the sum of the $\bar{\rho}_k^{Post}$ and $\bar{\rho}_k^{Pre}$ terms in (14), leading to a reduction in variance. Note that (14) reduces to the AR(1) estimator in Burlig et al. (2019) assuming no staggered treatment timing, $\sigma_\varepsilon^2 = 0$, and equal time intervals. The illustrative power analysis presented in Section 8 discusses other key features of (14).

Next, we present a corollary to Theorem 1 that provides new variance formulas for the point-in-time treatment effect estimators, $\hat{\beta}_{DID,l}^e$ and $\hat{\beta}_{DID,q}$.

**Corollary to Theorem 1.** The variance formula for $\hat{\beta}_{DID,l}^e$ in (9) that pertains to the DID estimator after $l$ periods of treatment exposure is

$$Var\left(\hat{\beta}_{DID,l}^e\right) = \frac{1}{(\sum_{k=1}^K I(l \leq A_k))^2} \sum_{k=1}^K I(l \leq A_k) \left(\frac{1}{M_{Tk}} + \frac{1}{M_{Ck}}\right) \left[\sigma_\theta^2 \left(1 + \frac{1}{B_k}\right.\right.$$
$$\left.\left. + \frac{(B_k - 1)}{B_k} \bar{\rho}_k^{Pre} - 2\bar{\rho}_{k,l}^{Pre-Post}\right) + \frac{\sigma_\varepsilon^2}{N}\right], \quad (15)$$

where $\bar{\rho}_{k,l}^{Pre-Post} = \frac{1}{B_k}\sum_{b=1}^{B_k} \rho^{(Time_{\{l+S_k-1\}} - Time_b)}$ is calculated at post-period $(l + S_k - 1)$.

Further, for the calendar-time DID estimator, we can calculate $Var(\hat{\beta}_{DID,q})$ using (15), replacing $I(l \leq A_k)$ with $I(q \geq S_k)$ and calculating $\bar{\rho}_{k,q}^{Pre-Post}$ at post-period $q$ rather than at post-period $(l + S_k - 1)$.

The result in (15) follows from Theorem 1 by averaging over only those timing groups with $l$ periods of observed exposure, setting $A_k = 1$, and translating exposure time into calendar time. The expression for $Var\left(\hat{\beta}_{DID,q}\right)$ can be obtained similarly.

In practice, standard errors for clustered DID designs using (7) are often estimated using OLS with cluster-robust standard errors (CRSE) using data aggregated to the cluster-time level (Cameron & Miller 2015; Liang & Zeger 1986). The CRSE approach assumes independence between clusters but allows for a general correlation structure within clusters. Based on simulation evidence in Bertrand et al. (2004), CRSE standard errors are typically adjusted for clustering at the unit of treatment assignment only.

*6.3. Cross-sectional Analysis: Power Calculations*

To facilitate power calculations, it is customary to express variance formulas in terms of ICCs of the error variances. We adopt this strategy by setting $\sigma_\theta^2 = ICC_\theta \sigma^2$ and $\sigma_\varepsilon^2 = (1 - ICC_\theta)\sigma^2$, where $\sigma^2 = \sigma_\theta^2 + \sigma_\varepsilon^2$ is the total error variance. Using this formulation, $\sigma^2$ no longer enters the $MDE$ formula in (1), and the power calculations can be conducted by specifying a value for $ICC_\theta$ rather than for the error variances directly.

Power calculations can now be conducted by inserting (14) or (15) into (1) to calculate $MDE$ values for pre-specified values of $P, M_{Tk}, M_{Ck}, N, ICC_\theta, A_k, B_k, \rho, \alpha$, and $\lambda$. Calculating the degrees of freedom $(df)$ for clustered designs is complex (Donald & Lang, 2007; Hedges, 2007). However, in panel settings, it is customary to use the number of cluster-time observations adjusted by the number of model parameters (Cameron & Miller 2015). Using (7), this approach yields $df = (MP - M - KP - \sum_k A_k)$ for the pooled estimator, $\hat{\beta}_{DID,POOL}$. In our running example, $df = 523$ (where $M = 79, P = 8, K = 3$, and $\sum_k A_k = 6$). Similarly, $df = (M_l P - $ $M_l - K_l P - K_l$) for the point-in-time estimator, $\hat{\beta}_{DID,l}^e$, where $M_l = \sum_k I(l \leq A_k)(M_{Tk} + M_{Ck})$ and $K_l = \sum_k I(l \leq A_k)$ are sample sizes for the included timing groups, and similarly for $\hat{\beta}_{DID,q}$.

Alternatively, (1) can be solved to calculate the total number of clusters, $M = M_T + M_C$, required to attain a given $MDE$ value. For example, for the pooled DID estimator, we can calculate the required cluster sample size using

$$M = \frac{Factor(\alpha, \lambda, df)^2 V}{MDE^2}, \tag{16}$$

where $V$ is the variance in (14), setting $M_{Tk} = r p_{Tk}$ and $M_{Ck} = (1-r) p_{Ck}$ using the inputs $r = M_T/M$, $p_{Tk} = M_{Tk}/M_T$, and $p_{Ck} = M_{Ck}/M_C$. Because $Factor(\alpha, \lambda, df)$ is also a function of $M$, iterative methods to solve nonlinear equations can be used to solve for $M$. For non-clustered designs, one can set $ICC_\theta$ equal to 0, $N$ to 1, and $M$ to the number of individuals.

### 6.4. Longitudinal Analysis

Under the longitudinal AR(1) design, the same individuals within the study clusters are followed over time. The analysis for this design extends the cross-sectional analysis, where the individual-level errors, $\varepsilon_{ijt}$, are now assumed to follow an AR(1) structure with autocorrelation parameter, $\psi$. Thus, under the longitudinal AR(1) design, we allow both cluster-level outcomes and individual-level outcomes within clusters to be correlated over time.

The resulting variance formulas are discussed in Appendix A.1. They parallel the structure for the cross-sectional estimators, except they now incorporate the added autocorrelation structure for the individual-level errors. For instance, with constant autocorrelations, we find that the variance for the longitudinal DID estimator for timing group $k$ is $Var(\hat{\beta}_{DID,k,long}) = \left(\frac{1}{M_{Tk}} + \frac{1}{M_{Ck}}\right)\left(\frac{1}{A_k} + \frac{1}{B_k}\right)[\sigma_\theta^2(1-\rho) + \frac{\sigma_\varepsilon^2}{N}(1-\psi)]$, where the $\frac{\sigma_\varepsilon^2}{N}$ term is now multiplied by $(1-\psi)$ to reflect outcome correlations of individuals over time. Note that setting $\sigma_\theta^2 = 0$ in this expression yields the variance estimator in Frison & Pocock (1992) and McKenzie (2012) for the non-clustered design. The same pattern holds for the AR(1) specification. In general, the longitudinal estimators will have less precision than the cross-sectional estimators, except for special cases that parallel those discussed in Section 6.2 above for the cross-sectional analysis.

*6.5. Incorporating Covariates*

In DID analyses, a vector of time-varying covariates, $\mathbf{x}_{ijt}$, with associated parameter vector, $\boldsymbol{\gamma}$, is often included in the models to adjust for potential confounding bias. To examine the effects of covariates on precision, to fix concepts, we begin by assuming no confounding bias by invoking a parallel trends assumption for the covariates (which we then relax):

**Assumption DID.2. Parallel trends for the covariates.** For each timing group ($k \geq 1$), post-treatment time period ($q \geq S_k$), and covariate $c \in (1, \dots, v)$:

$$E\left(\bar{x}_{c,\{t=q\}} - \bar{\bar{x}}_{c,\{t<S_k\}} \middle| G_{Tj} = k\right) = E\left(\bar{x}_{c,\{t=q\}} - \bar{\bar{x}}_{c,\{t<S_k\}} \middle| G_{Cj} = k\right), \tag{17}$$

where $v$ is the number of covariates and the $\bar{x}_{c,\{t=q\}}$ and $\bar{\bar{x}}_{c,\{t<S_k\}}$ covariate means are defined analogously to the $\bar{Y}(0)$ and $\bar{\bar{Y}}(0)$ outcome means defined in (2) and (6) above.

Because $\boldsymbol{\gamma}$ is assumed constant over time and across timing groups, the implication of this assumption is that mean covariate values during the pre- and post-treatment periods are independent of treatment status for each timing group. The result is that the addition of covariates to the models does not change the DID estimators in expectation.

To see this result more formally, it can be shown that OLS estimators for $\beta_{DID,kq}$ with and without covariates, $\hat{\beta}_{DID,kq,X}$ and $\hat{\beta}_{DID,kq}$, can be related using $\hat{\beta}_{DID,kq,X} = \hat{\beta}_{DID,kq} - \left([\bar{\mathbf{x}}_{\{t=q\}}^{G_{Tj}=k} - \bar{\bar{\mathbf{x}}}_{\{t<S_k\}}^{G_{Tj}=k}] - [\bar{\mathbf{x}}_{\{t=q\}}^{G_{Cj}=k} - \bar{\bar{\mathbf{x}}}_{\{t<S_k\}}^{G_{Cj}=k}]\right)' \hat{\boldsymbol{\gamma}}$, where $\bar{\mathbf{x}}$ and $\bar{\bar{\mathbf{x}}}$ are vectors of covariate means, and $\hat{\boldsymbol{\gamma}}$ is the OLS estimator for $\boldsymbol{\gamma}$ (see also Schochet et al., 2021). Note that the term in parentheses is a vector containing DID estimators for each covariate. But these DID estimators each have zero expectation under Assumption DID.2. Thus, $E(\hat{\beta}_{DID,kq,X}) = E(\hat{\beta}_{DID,kq})$, even when $\hat{\gamma} \neq 0$.

Thus, under Assumption DID.2, precision gains from covariates can be quantified by multiplying the variance terms from the model without covariates by $(1 - R_{YX}^2)$, where $R_{YX}^2$ is the $R^2$ value from covariate inclusion. We assume the same $R^2$ value to explain $\sigma_\theta^2$ and $\sigma_\varepsilon^2$. With covariates, $df$ is reduced by the number of covariates, $v$. This approach parallels the power analysis approach for RCTs where baseline covariates are independent of treatment status due to randomization (Raudenbush, 1997; Schochet, 2008).

Assumption DID.2, however, is not likely to hold in practice. Thus, precision gains from covariates will typically be reduced by treatment-covariate collinearity. Thus, the net variance reduction due to the covariates can be approximated using the ratio $(1 - R_{YX}^2)/(1 - R_{TX}^2)$, where $R_{TX}^2$ is the average $R^2$ value from regressing the treatment interaction terms in (7) on the covariates. This ratio can be less than 1. We summarize this result in the following theorem.

**Theorem 2.** If covariates are added to the model in (7), the variance formula for $\hat{\beta}_{DID,POOL,X}$ can be approximated using

$$Var(\hat{\beta}_{DID,POOL,X}) = \frac{(1 - R_{YX}^2)}{(1 - R_{TX}^2)} Var(\hat{\beta}_{DID,POOL}), \tag{18}$$

and similarly for the point-in-time estimators, $\hat{\beta}_{DID,l}^e$ and $\hat{\beta}_{DID,q}$, and the longitudinal estimators.

### 7. Comparative Interrupted Time Series (CITS) and ITS Estimators

CITS estimators are based on specifications that model trends before and after the introduction of the treatment. They quantify whether once the treatment begins, the treatment group deviates from its pre-intervention trend by a greater amount than does the comparison group. The more common ITS estimators pertain to designs without a comparison group, where the treatment group is compared to its own pre-period trend only. CITS and ITS estimators were popularized in the seminal works of Cook and Campbell (1979) and Shadish et al. (2002); recent applications and reviews in various disciplines are provided in Bloom (2003), Somers et al. (2013), Kontopantelis et al. (2015), Linden et al. (2015), St. Clair et al. (2016), Bernal et al. (2017), and Baicker and Svoronos (2019). We assume at least 3 (preferably 4) pre- and post-period time points each so that the trends can be adequately modeled (Cook & Campbell, 1979).

For our CITS and ITS power analysis, we adopt an event history approach for the reasons discussed above for the DID design. We model pre- and post-period trends using a linear specification, where we allow the intercepts and slopes to differ across the two time periods, which we refer to as the "fully-interacted" model. This approach can be extended to allow for segmented regression lines across different post-periods. We also present results for a restricted model that assumes common slopes across the pre- and post-periods. Bernal et al. (2017) and Ferron and Rendina-Gobioff (2014) discuss other CITS estimands not considered here.

We rely on the following fully-interacted regression model using data for all timing groups:

$$y_{ijkt} = \alpha_{0k}T_{jk}Pre_{kt} + \gamma_{0k}Time_t T_{jk}Pre_{kt} + \alpha_{1k}T_{jk}Post_{kt} + \gamma_{1k}Time_t T_{jk}Post_{kt}$$
$$+ \delta_{0k}(1 - T_{jk})Pre_{kt} + \tau_{0k}Time_t(1 - T_{jk})Pre_{kt} + \delta_{1k}(1 - T_{jk})Post_{kt} \quad (19)$$
$$+ \tau_{1k}Time_t(1 - T_{jk})Post_{kt} + \theta_{jkt} + \varepsilon_{ijkt},$$

where $Pre_{kt}$ is an indicator that equals 1 if the time period corresponds to the pre-period (that is, if $t < S_k$) and 0 otherwise, and $Post_{kt} = 1 - Pre_{kt}$ is an indicator that corresponds to the post-period, recalling that $Time_t$ denotes when the outcome was measured in calendar time relative to a common reference point.

In this model, $\alpha_{0k}$ and $\gamma_{0k}$ are pre-period intercept and slope parameters for treatments in timing group $k$; $\alpha_{1k}$ and $\gamma_{1k}$ are post-period intercepts and slopes for treatments in timing group $k$; and $\delta_{0k}$, $\tau_{0k}$, $\delta_{1k}$, and $\tau_{1k}$ are corresponding parameters for the matched comparison group. In this formulation, we include the treatment-by-time indicator terms (such as $T_{jk}Pre_{kt}$) rather than the cluster-level fixed effects as we did for the DID design to simplify variance estimation (both specifications yield the same parameter estimates). Using our running example, the model in (19) contains 24 parameters (8 pre- and post-period intercept and slope parameters for each of the three timing groups, split evenly between the treatment and comparison groups).

Applying OLS to the model in (19), we can estimate the $ATT_{kq}$ parameter in (2) using

$$\hat{\beta}_{CITS,kq} = \left([\hat{\alpha}_{1k} + \hat{\gamma}_{1k}Time_q] - [\hat{\alpha}_{0k} + \hat{\gamma}_{0k}Time_q]\right)$$
$$- \left([\hat{\delta}_{1k} + \hat{\tau}_{1k}Time_q] - [\hat{\delta}_{0k} + \hat{\tau}_{0k}Time_q]\right) \quad (20)$$
$$= \left(Pred1_{kq} - Pred2_{kq}\right) - \left(Pred3_{kq} - Pred4_{kq}\right).$$

The first part of this estimator, $(Pred1_{kq} - Pred2_{kq})$, measures the difference between the predicted values at time $q$ based on the fitted pre- and post-period trendlines for the treatment group, and the second part, $(Pred3_{kq} - Pred4_{kq})$, measures the pre-post difference—the "forecast error"—for the matched comparison group. Note that $(Pred2_{kq} + Pred3_{kq} - Pred4_{kq})$ is the estimated counterfactual outcome at time $q$ for the treatment group in the absence of the intervention; it incorporates both the forecast at time $q$ from the fitted treatment group pre-period trendline as well as the comparison group forecast error.

Using (20), we see that the CITS estimator differs from the DID estimator in that it measures treatment-comparison differences in estimated trendlines, not just estimated intercepts; thus, the CITS estimator will have a larger variance (which we formalize below). The ITS estimator, $\hat{\beta}_{ITS,kq} = (Pred1_{kq} - Pred2_{kq})$, is based on the treatment group trendlines only.

As with the DID approach, we can now average the $\hat{\beta}_{CITS,kq}$ estimators over $k$ to yield the point-in-time CITS estimators, $\hat{\beta}_{CITS,q}$ and $\hat{\beta}^e_{CITS,l}$, which can then be averaged over their post-periods to yield the pooled estimator, $\hat{\beta}_{CITS,POOL}$ (and similarly for the ITS estimators). These estimators are unbiased for the $ATT_q$, $ATT^e_l$, and $ATT_{POOL}$ parameters under Assumptions 1 and 2, assuming the linear specification in (19) is correct, and under the following added assumption:

**Assumption CITS.1. Parallel deviations from pre-period trendlines.** In the absence of the intervention, deviations from pre-treatment trendlines would be equivalent for the treatment and comparison groups for all timing groups ($k \geq 1$) and post-periods ($q \geq S_k$). This assumption can be expressed in terms of potential outcomes in the untreated condition as

$$E\big(\bar{Y}_{\{t=q\}}(0) - Pred2_{kq}\big|G_{Tj} = k\big) = E\big(\bar{Y}_{\{t=q\}}(0) - Pred4_{kq}\big|G_{Cj} = k\big). \tag{21}$$

This condition requires that in the absence of treatment, forecast errors from the pre-period trendlines would be the same, on average, for the treatment and comparison groups. The parallel condition for the ITS estimator is perfect mean forecasts for the treatment group:

$E\big(\bar{Y}_{\{t=q\}}(0) - Pred2_{kq}\big|G_{Tj} = k\big) = 0$.

To calculate the variance of $\hat{\beta}_{CITS,kq}$, we can use (20) to compute the variances and covariances of the $Pred_{kq}$ terms. To facilitate the calculations, note that the OLS estimates from (19) are identical to those from separate models estimated using four groups of observations for each timing group $k$: Group 1: treatments in the post-period (to obtain $Pred1_{kq}$); Group 2: treatments in the pre-period (to obtain $Pred2_{kq}$); Group 3: comparisons in the post-period (to obtain $Pred3_{kq}$); and Group 4: comparisons in the pre-period (to obtain $Pred4_{kq}$).

For illustration, consider OLS estimation using Group 2 data for timing group $k$ (with $M_{Tk}$ clusters) aggregated to the cluster-time level for a model with an intercept and time variable. For estimation, we center $Time_t$ around its pre-period mean, $\overline{Time}_{\{t<S_k\}}$, which facilitates the calculations, because $\hat{\alpha}_{0k}$ now becomes the pre-period mean outcome, $\bar{\bar{y}}_{\{t<S_k\}}^{G_{Tj}=k}$.

Consider first the model without the AR(1) error structure that we use to highlight key features of the more general, complex variance estimator presented in Theorem 3 below. For this specification, we find after applying standard OLS methods that the variance of the $Pred2_{kq}$ forecast at time $q$ based on the fitted pre-period trendline for Group 2 is

$$Var(Pred2_{kq}) = \frac{1}{M_{Tk}}\left(\frac{1}{B_k} + \frac{(Time_q - \overline{Time}_{\{t<S_k\}})^2}{SSQT_k^{Pre}}\right)\left(\sigma_\theta^2 + \frac{\sigma_\varepsilon^2}{N}\right), \tag{22}$$

where $SSQT_k^{Pre} = \sum_{b=1}^{B_k}(Time_b - \overline{Time}_{\{t<S_k\}})^2$ is the pre-period variance of the time variable.

In this expression, the terms inside the first brackets capture the variances of the estimated intercept and slope parameters as well as their covariance. We see that precision decreases for later post-periods than earlier ones (as $q$ increases) because the forecasts become more tentative (as reflected in a larger value for the $(Time_q - \overline{Time}_{\{t<S_k\}})^2$ term). Further, increasing the number of pre-periods, $B_k$, will increase precision both because $(1/B_k)$ becomes smaller and $SSQT_k^{Pre}$ becomes larger as the trendlines become more precisely estimated. Similar variance expressions exist for the $Pred1_{kq}$, $Pred3_{kq}$, and $Pred4_{kq}$ terms.

Because the four $Pred_{kq}$ groups are independent in the model without the AR(1) error structure, we can sum their variances to obtain the following variance expression for $\hat{\beta}_{CITS,kq}$ for the model without autocorrelated errors:

$$Var(\hat{\beta}_{CITS,kq}) = \left(\frac{1}{M_{Tk}} + \frac{1}{M_{Ck}}\right)\left(\frac{1}{A_k} + \frac{1}{B_k} + \frac{(Time_q - \overline{Time}_{\{t \geq S_k\}})^2}{SSQT_k^{Post}} \right.$$
$$\left. + \frac{(Time_q - \overline{Time}_{\{t<S_k\}})^2}{SSQT_k^{Pre}}\right)\left(\sigma_\theta^2 + \frac{\sigma_\varepsilon^2}{N}\right), \tag{23}$$

where $SSQT_k^{Post} = \sum_{a=S_k}^{P}(Time_a - \overline{Time}_{\{t \geq S_k\}})^2$ is the post-period variance of the time variable, and other terms are defined above. The parallel variance expression for the ITS estimator omits the $(1/M_{Ck})$ term in (23).

Comparing (23) to the variance in (13) for the corresponding DID estimator, we see that the CITS variance is *larger* than the DID variance because of the addition of the two time-related terms that reflect the estimation error in the fitted trendlines. Thus, the CITS estimator will require larger sample sizes than the DID estimator to yield impacts with the same precision level. The DID power gains come at the expense of stronger conditions for parameter identification.

Note that the regression $R^2$ value from the model in (19)—that measures the fit of the trendlines— enters the variance formula in (22) through its effects on the error variances, $\sigma_\theta^2$ and $\sigma_\varepsilon^2$. Thus, a stronger linear time-outcome relationship leads to smaller values of $\sigma_\theta^2$ and $\sigma_\varepsilon^2$. But these $R^2$ values do not enter the power calculations—as measured in effect size units—because they also enter the error variances in the denominator of (1) that are used to construct the $MDE$ values, and thus cancel. Similar issues apply to the DID estimator.

As with the DID estimator, including the AR(1) structure considerably complicates the CITS analysis due to the emergence of correlations between the fitted pre- and post-period trendlines. To calculate the variance formulas in this setting, consider, as before, estimating $Pred2_{kq}$ using Group 2 data aggregated to the cluster-time level with $M_{Tk}B_k$ observations. Switching to matrix notation to simplify notation, let $\mathbf{Z}_{2k}$ be the matrix of independent variables from the regression model that includes the constant term and centered time variables, with parameter vector, $\boldsymbol{\eta}_{2k} = (\alpha_{0k}\ \gamma_{0k})'$. With the AR(1) error structure, the variance of the OLS estimator, $\widehat{\boldsymbol{\eta}}_{2k}$, becomes $Var(\widehat{\boldsymbol{\eta}}_{2k}) = (\mathbf{Z}'_{2k}\mathbf{Z}_{2k})^{-1}(\mathbf{Z}'_{2k}\boldsymbol{\Omega}_k\mathbf{Z}_{2k})(\mathbf{Z}'_{2k}\mathbf{Z}_{2k})^{-1}$, where $\boldsymbol{\Omega}_k$ is an $M_{Tk}B_k \times M_{Tk}B_k$ block-diagonal error variance-covariance matrix, containing the $B_k \times B_k$ block sub-matrices, $\boldsymbol{\Omega}_{jk}$, for the $B_k$ observations in the same cluster, with entries $\left(\sigma_\theta^2 + \frac{\sigma_\varepsilon^2}{N}\right)$ along the diagonal and $\sigma_\theta^2 \rho^{|Time_a - Time_b|}$ along the $(a,b)$ off-diagonal cells.

Using this framework, we can then compute $Var(Pred2_{kq}) = Var(\hat{\alpha}_{0k}) + Time_q^2 Var(\hat{\gamma}_{0k}) + 2Time_q Cov(\hat{\alpha}_{0k}, \hat{\gamma}_{0k})$, and similarly for the variances of $Pred1_{kq}$, $Pred3_{kq}$, and $Pred4_{kq}$. If we then calculate the covariances across these estimators and then average across post-periods and timing groups, we obtain a complex, but closed-form expression for the variance of the pooled CITS estimator, $\hat{\beta}_{CITS,POOL}$, that we present in the following new theorem.

**Theorem 3.** The variance of the pooled CITS estimator, $\hat{\beta}_{CITS,POOL}$, obtained from the model in (19) that incorporates clustering, variation in intervention timing, the AR(1) error structure, and general measurement intervals is as follows:

$$Var(\hat{\beta}_{CITS,POOL})$$
$$= Var(\hat{\beta}_{DID,POOL})$$
$$+ \frac{1}{(\sum_{k=1}^K A_k)^2} \sum_{k=1}^K A_k^2 \left(\frac{1}{M_{Tk}}\right. \tag{24}$$
$$\left. + \frac{1}{M_{Ck}}\right) \left[\sigma_\theta^2 (Term1_k + Term2_k - Term3_k) + \left(\frac{\sigma_\varepsilon^2}{N}\right) Term4_k\right],$$

where

$$Term1_k = \left(\overline{Time}_{\{t \geq S_k\}} - \overline{Time}_{\{t < S_k\}}\right)^2 \left(\frac{1}{SSQT_k^{Pre}} + \frac{(B_k - 1)B_k \bar{\rho}_k^{Pre1}}{[SSQT_k^{Pre}]^2}\right),$$

$$Term2_k = 2\left(\overline{Time}_{\{t \geq S_k\}} - \overline{Time}_{\{t < S_k\}}\right) B_k \bar{\rho}_k^{Pre2} \left(\frac{1}{SSQT_k^{Pre}}\right),$$

$$Term3_k = 2\left(\overline{Time}_{\{t \geq S_k\}} - \overline{Time}_{\{t < S_k\}}\right) B_k \bar{\rho}_k^{Pre-Post1} \left(\frac{1}{SSQT_k^{Pre}}\right),$$

$$Term4_k = \left(\overline{Time}_{\{t \geq S_k\}} - \overline{Time}_{\{t < S_k\}}\right)^2 \left(\frac{1}{SSQT_k^{Pre}}\right),$$

$$\bar{\rho}_k^{Pre1} = \frac{2}{B_k(B_k-1)} \sum_{b=1}^{B_k} \sum_{b'>b}^{B_k} (Time_b - \overline{Time}_{\{t<S_k\}})(Time_{b'} - \overline{Time}_{\{t<S_k\}}) \rho^{(Time_{b'}-Time_b)},$$

$$\bar{\rho}_k^{Pre2} = \frac{1}{B_k^2} \sum_{b=1}^{B_k} \sum_{b'=1}^{B_k} (Time_b - \overline{Time}_{\{t<S_k\}}) \rho^{|Time_{b'}-Time_b|},$$

$$\bar{\rho}_k^{Pre-Post1} = \frac{1}{A_k B_k} \sum_{b=1}^{B_k} \sum_{q=S_k}^{P} (Time_b - \overline{Time}_{\{t<S_k\}}) \rho^{(Time_q-Time_b)},$$

and other terms are defined above. This variance formula can be used for the power calculations with $df = (MP - 8K)$. The ITS variance formula omits the $(1/M_{Ck})$ term in (24).

The $Term_k$ expressions in (24) have a ready interpretation. $Term1_k$ is the average variance of the forecasts from the fitted pre-period regression lines for both the treatment and comparison groups, where averaging is taken over all post-periods. $Term2_k$ captures the covariances between the estimated intercepts and slopes from the fitted pre-period regression lines (which are zero for evenly spaced time periods or for constant autocorrelations). $Term3_k$ pertains to the covariances between the estimated intercepts from the fitted pre-period trendlines and the slopes from the fitted post-period trendlines. In these expressions, the $\bar{\rho}$ terms arise due to the $(\mathbf{Z'\Omega Z})$ matrices from above and pertain to various time-related variances weighted by the autocorrelations. Finally, $Term4_k$ is similar to $Term1_k$ but pertains to the individual-level errors which are uncorrelated over time, so it does not include the $\bar{\rho}$ terms.

It is interesting that for the pooled estimator, $\hat{\beta}_{CITS,POOL}$, the predicted values from the post-period trendlines averaged over all post-periods are simply the mean post-period outcomes for the treatments and comparisons, $\bar{\bar{y}}_{\{t \geq S_k\}}^{G_{Tj}=k}$ and $\bar{\bar{y}}_{\{t \geq S_k\}}^{G_{Cj}=k}$. Thus, the variance in (24) does not account for the estimation error in the post-period slopes. For the same reason, the variance in (24) also applies to a variant of the model in (19) where the post-period is modeled using *discrete* post-period indicators rather than using a linear trendline (see, e.g., Bloom, 1999, Somers et al., 2013 and St. Clair et al., 2016 that use this specification).[5]

The corollary to Theorem 3 in Appendix A.2 shows the associated variances for the point-in-time CITS estimators, $\hat{\beta}^e_{CITS,l}$ and $\hat{\beta}_{CITS,q}$. It is surprising that these variances are more complex than for the pooled CITS estimator, as they must also account for the estimation error in the post-period slopes. In these variances, if we set $\rho = 0$ and assume one timing group ($K = 1$), then at calendar time $q$, we have that $Var(\hat{\beta}_{CITS,q}) = \left(\frac{1}{M_T} + \frac{1}{M_C}\right)\left(1 + \frac{1}{B} + \frac{(Time_q - \overline{Time}_{\{t<S\}})^2}{SSQT^{Pre}}\right)(\sigma_\theta^2 + \frac{\sigma_\varepsilon^2}{N})$, which is the expression in Bloom (1999). For the $MDE$ calculations, we can use $df = (M_l P - 8K_l)$ for the $\hat{\beta}^e_{CITS,l}$ estimator, where $M_l = \sum_k I(l \leq A_k)(M_{Tk} + M_{Ck})$ and $K_l = \sum_k I(l \leq A_k)$ are sample sizes for the included timing groups, and similarly for $\hat{\beta}_{CITS,q}$.[6]

Next, we consider a popular restricted variant of the model in (19) that assumes common slope parameters for the pre- and post-period trendlines, where we omit the $Time_t T_{jk} Pre_{kt}$ and $Time_t (1 - T_{jk}) Pre_{kt}$ terms. For this "common-slopes" specification, the OLS estimators of interest are $\hat{\beta}_{CITS,CS,k} = (\hat{\alpha}_{1k} - \hat{\alpha}_{0k}) - (\hat{\delta}_{1k} - \hat{\delta}_{0k})$, which are treatment-comparison differences between the estimated pre- and post-period intercepts. For this specification, the pooled and point-in-time treatment effects are the same because this approach assumes a constant treatment effect over time (for each $k$).

To obtain variance expressions for $\hat{\beta}_{CITS,CS,k}$, we can use the same methods as for the fully-interacted estimators above. Without the AR(1) structure, we find that

$$Var(\hat{\beta}_{CITS,CS,k}) = \left(\frac{1}{M_{Tk}} + \frac{1}{M_{Ck}}\right)\left(\frac{1}{A_k} + \frac{1}{B_k}\right)\left(\frac{SSQT^{Full}}{SSQT_k^{Pre} + SSQT_k^{Post}}\right)\left(\sigma_\theta^2 + \frac{\sigma_\varepsilon^2}{N}\right), \quad (25)$$

where $SSQT^{Full} = \sum_{p=1}^{P}(Time_p - \overline{Time}_{\{t\geq 1\}})^2$ is the variance of the time variable computed over the full observation period, and other terms are defined above. We see that this variance gets smaller as $B_k$ or $A_k$ increases and is symmetric if $B_k$ and $A_k$ are switched (if time measurements are even). Further, comparing (25) to (13), we see that the variance of the common-slopes estimator is *larger* than for the DID estimator, because the third bracketed term in (25) is greater than 1. Further, our illustrative power analysis in Section 8 shows that the variance of the common-slopes estimator is considerably smaller than for the fully-interacted pooled CITS estimator which has weaker parameter identification assumptions.

---

[5] However, the degrees of freedom for this discrete specification is different with $df = (MP - 4K - \sum_{k=1}^{K} A_k)$.

[6] For the CITS specification with discrete post-period indicators, the point-in-time estimators can be obtained using (24) evaluated at post-period $q$ and setting $A_k = 1$.

Theorem 4 in Appendix A.3 presents the more complex variance formula for the common-slopes estimator that includes the AR(1) structure. For power calculations using this estimator, we can use $df = (MP - 6K)$. Clearly, the choice of the common-slopes or fully-interacted specification is an empirical issue and can be tested by examining the statistical significance of the differences in the estimated pre- and post-period trendlines.

Issues regarding the inclusion of model covariates are similar for the CITS and ITS estimators as for the DID estimators. For instance, invoking the parallel deviations assumption in (21) *for each covariate*, the variance terms in the model without covariates can be multiplied by $(1 - R_{YX}^2)$. This holds because the CITS estimator with and without model covariates, $\hat{\beta}_{CITS,kq,X}$ and $\hat{\beta}_{CITS,kq}$, can be related using $\hat{\beta}_{CITS,kq,X} = \hat{\beta}_{CITS,kq} - \hat{\varphi}'_{CITS,kq}\hat{\gamma}$, where $\hat{\gamma}$ are parameter estimates for the covariates and $\hat{\varphi}_{CITS,kq}$ is a vector of CITS estimators for the covariates when each covariate is sequentially treated as the dependent variable in the model in (19). Under the parallel deviations assumption for each covariate, $\hat{\varphi}_{CITS,kq}$ has zero expectation, so $E(\hat{\beta}_{CITS,kq,X}) = E(\hat{\beta}_{CITS,kq})$. However, in practice, it is more realistic to multiply the variance terms by the factor $(1 - R_{YX}^2)/(1 - R_{TX}^2)$ to adjust for treatment-covariate collinearity.

Finally, for the AR(1) longitudinal design, the off-diagonal block entries of $\mathbf{\Omega}_{jk}$ become $(\sigma_\theta^2 \rho^{|Time_a - Time_b|} + \frac{\sigma_\varepsilon^2}{N}\psi^{|Time_a - Time_b|})$, which yields a parallel structure for the cluster- and individual-level variance components, leading to parallel variance formulas (see Appendix A.1).

## 8. An Illustrative Power Analysis

To highlight key features of the variance formulas for the considered DID, CITS, and ITS estimators, this section presents an illustrative power analysis. Due to the large number of parameters that enter the analysis, our goal is not to present a compendium of power results across many combinations of parameter values. Rather, we broadly address two power-related questions that align with the focus of our theoretical analysis: (1) How does variation in treatment timing and the AR(1) error structure affect precision? and (2) To what extent does precision differ for the DID, CITS, and ITS estimators? We conduct our analysis using plausible parameter values, with a focus on sample size requirements to attain a given $MDE$ value for a two-tailed significance test ($\alpha = 0.05$) at 80 percent power ($\lambda = 0.8$) using the power formula in (16). The calculations were conducted using the *Power_Panel* dashboard that was developed for this article (add link) that readers can use to conduct power analyses for their specific designs.

Our first main finding is that *both variation in treatment timing and the AR(1) error structure increase required cluster sample sizes*. These increases become more pronounced if there are a small number of pre-periods for *any* timing group (which occurs in our setting when $S_1$ and $S_2$ diverge) and as the AR(1) autocorrelation coefficient ($\rho$) increases (but only up to a point).

Figure 1 illustrates these findings using the pooled DID estimator. The figure shows design effects—the ratio of sample size requirements ($M$) for a design with staggered treatment timing and/or an AR(1) error structure relative to a reference design without these design features (or that ignores them, which yields bias). The calculations assume $P = 8$ time periods (as is the case in our running example), a cross-sectional design, equal time spacing of measurements, a 50-50 treatment-comparison split, $N = 100$ individuals per cluster-time cell, no model covariates, and an $ICC_\theta$ of 0.05—a common value found in education research (Hedges & Hedberg, 2007; Schochet, 2008). For the design with staggered treatment timing, we assume $K = 2$ timing groups of equal size with various start times ($S_1$ and $S_2$). For the reference design with $K = 1$, we set the treatment start date ($S$) as the average of $S_1$ and $S_2$, rounded to the nearest integer. The design effects were then calculated by varying $\rho$, $S_1$, and $S_2$. Note that the results hold for any $MDE$ value, because this value cancels when calculating the design effects using (16).

As shown in Figure 1, the presence of both staggered treatment timing and the AR(1) error structure yields design effects for the pooled DID estimator that range from 1.13 to 2.32 across the considered designs, with a mean of 2.05, or a doubling of required cluster samples (Panel 1a). Staggered treatment timing matters as can be seen by examining the design effects when $\rho = 0$, which become larger as $S_1$ and $S_2$ become more spread out (that is, as timing group 1 has fewer pre-periods). The AR(1) structure also matters as evidenced by increases in the design effects as $\rho$ increases, until $\rho$ gets large or if the pre- or post-periods are short, as predicted by the theory from Section 6 (Panel 1b). Similar results hold for the CITS and ITS estimators (not shown).

Our second main finding is that *the CITS estimators require larger cluster samples than the DID estimator to attain the same level of precision, especially for the fully-interacted CITS estimator.* This occurs due to the estimation error in the fitted CITS (and ITS) trendlines. The effects are largest for the fully-interacted CITS estimator, which estimates four trendlines compared to only two for the common-slopes CITS estimator and half as many for the ITS estimators (which exclude the comparison group).

These results are illustrated in Table 3, which shows cluster sample sizes ($M$) required to achieve an $MDE$ of 0.20—a common target used for clustered education RCTs (see, e.g., Schochet, 2008). We present results for the pooled estimators averaged across all post-periods as well as for the point-in-time estimators measured 1, 3, and 5 periods after treatment exposure. We apply many of the same assumptions as for Figure 1, except we now consider a broader range of time periods ($P$) and treatment start times ($S_1$ and $S_2$) to allow for more pre- and post-periods for the CITS and ITS estimators (recall that the fully-interacted estimators require at least 3 pre- and 3 post-periods each) and we fix $\rho$ at 0.4 (based on our NAEP analysis presented in Section 6.1). We also present selected results for the pooled estimators for the model with constant autocorrelations and for the longitudinal AR(1) design (with $\psi = 0.4$).

Consider first the top panel of Table 3 showing results for the *pooled* estimators. We see that for all designs, the fully-interacted CITS pooled estimator (which is identical to the discrete CITS pooled estimator) requires larger cluster samples than the other estimators. For instance, when $P = 8$, $S_1 = 4$, and $S_2 = 6$, the fully-interacted CITS estimator requires $M = 297$ total clusters, compared to $M = 37$ for the DID estimator and $M = 89$ for the common-slopes CITS estimator (the ITS estimators require half as many treatment group clusters as their associated CITS estimators). Further, if $P$ increases to 12, then with $S_1 = 4$ and $S_2 = 8$, the required sample size balloons to $M = 641$ for the fully-interacted CITS estimator, but decreases to $M = 32$ for the DID estimator and to $M = 68$ for the common-slopes CITS estimator.

Sample sizes for the fully-interacted pooled estimators increase with $P$ (all else equal) because their point-in-time forecast errors increase considerably over time (see below). Further, for a given $P$, these estimators tend to lose precision with fewer pre-periods ($B_k$) and more post-periods ($A_k$). In contrast, the pooled DID and common-slopes estimators do not exhibit these patterns; rather their precision tends to increase as $P$ increases and are less sensitive to specific values of $B_k$ and $A_k$ (holding $P$ fixed). Nonetheless, the pooled common-slopes estimator requires 2 to 3 larger cluster samples than the pooled DID estimator (Table 3). Note that precision for the DID estimator noticeably improves if $B_k$ increases from 1 to 3.

The top panel of Table 3 also shows that consistent with theory, required sample sizes reduce for the model with constant autocorrelations, as this specification actually improves power. These reductions are largest for the DID pooled estimator, where required samples are less than half of those needed for the model with AR(1) errors. We find also that sample sizes are only slightly larger for the pooled longitudinal AR(1) estimator than the pooled cross-sectional one.

The bottom three panels of Table 3 provide additional perspective on the pooled findings using the *point-in-time estimators*. First, we see that if $B_k \geq 5$ for all $k$, power levels for the fully-interacted and common-slopes CITS estimators are comparable for one-period forecasts, although they both still have less power than the DID estimator. In this setting, the ITS estimators require fewer treatment group clusters than the DID estimator. However, precision levels for the fully-interacted estimators rapidly decay for the three- and five-period forecasts (as is also the case for the discrete estimators [not shown]). For example, when $P = 12$, $S_1 = 6$, and $S_2 = 8$, the required sample size for the fully-interacted CITS estimator is $M = 74$ for the one-period forecast, compared to $M = 127$ for the three-period forecast and $M = 250$ for the five-period forecast. In contrast, while the DID estimator loses some precision over time (because the $\bar{\rho}_{k.l}^{Pre-Post}$ term in (15) decreases), these effects are small—apart from some five-period forecasts that have low power because they are calculated excluding timing group 2 for whom period 5 is not an observed post-period (e.g., when $P = 12$, $S_1 = 6$, and $S_2 = 10$). Similarly, the common-slopes estimator maintains its precision levels over time as it assumes constant treatment effects (apart from the five-period forecasts that exclude timing group 2).

The same pattern of results holds if we vary other power parameters, although the required $M$ values can change. For instance, if the targeted $MDE$ value is halved from 0.20 to 0.10, the required $M$ increases fourfold relative to those shown in Table 3 for all estimators (because $M$ is inversely proportional to $MDE^2$ as shown in (16)). Similarly, the required $M$ is roughly proportional to the $ICC_\theta$ value (for moderate $N$) and inversely proportional to $(1 - R_{YX}^2)/(1 - R_{TX}^2)$ for models with covariates (see Theorem 2). Further, precision tends to decrease with more timing groups. The results are less sensitive to the number of individuals per cluster-time cell, $N$, as precision for clustered designs is typically driven by $M$, unless $N$ is small. For instance, when $P = 8$, $S_1 = 4$, and $S_2 = 6$ the required $M$ for the pooled common-slopes CITS estimator is $M = 89$ for $N = 100$, compared to $M = 75$ for $N = 1{,}000$ and $M = 103$ for $N = 50$. The results are also relatively insensitive to the $M_{Tk}$ and $M_{Ck}$ allocations as long as they range from 0.3 to 0.7.

## 9. Conclusions

This article developed new closed-form variance expressions for power analyses for commonly used DID, CITS, and ITS regression estimators. The main contribution was to incorporate variation in treatment timing into the variance formulas, but the formulas also account for several other key design features that arise in practice: autocorrelated errors, unequal measurement intervals, and clustering due to the unit of treatment assignment. Using an event history approach, we considered power formulas for both cross-sectional and longitudinal data structures and allowed for the inclusion of model covariates that can improve precision. Further, we considered both point-in-time estimators as well as those pooled over the entire post-period.

These variance formulas can be used to calculate $MDE$ values for a given number of time periods and study clusters or, conversely, to calculate required sample sizes to attain a given $MDE$. The free *Panel_Power* dashboard (add link) can be used to perform the sample size calculations.

Our theory and illustrative power analysis both demonstrated that variation in treatment timing and the AR(1) error structure increase required cluster sample sizes for the considered panel estimators. Thus, it is important that these design features (if pertinent) be considered when assessing appropriate sample sizes in designing panel studies.

Further, our analysis showed that the CITS estimators require larger cluster samples than the DID estimator to attain the same level of precision, especially for the fully-interacted and discrete CITS estimators. The reason is simple: the CITS and ITS estimators must account for the estimation error in the fitted trendlines, not just the fitted intercepts. These effects are most pronounced for the fully-interacted CITS estimator (which estimates four trendlines), but are also present for the common-slopes CITS estimator (which estimates two trendlines) and the corresponding ITS estimators. Power for the fully-interacted and discrete point-in-time estimators deteriorate rapidly over time, yielding large sample size requirements for the pooled estimators averaged over multiple post-periods. While these time effects do not affect the common-slopes CITS estimator, it still requires about 2 to 3 larger cluster samples than the pooled DID estimator across the considered designs. Power losses for the CITS and ITS estimators would be even greater assuming more complex specifications of the time variable in the model (e.g., quadratic relationships) and more complex error structures.

The results suggest a tradeoff between estimator precision and the strength of the identification assumptions for obtaining unbiased ATT estimators. The DID estimator requires the smallest sample sizes—that may be attainable in practice, especially if there are at least three pre-period observations—but imposes the strongest identification conditions. In contrast, the fully-interacted CITS estimator imposes weaker identification conditions but requires large samples (except for the point-in-time estimators soon after treatment exposure if there are at least five pre-period observations). The common-slopes CITS estimator lies somewhere in between. The ITS estimators have more power than the CITS estimators as they avoid the variance contribution from the comparison group, but typically have less power than the DID estimator. Thus, while the fully-interacted CITS estimator may best control potential selection biases, it may be less feasible for many studies from a power (and mean squared error) perspective. In these cases, the DID, ITS, and common-slopes estimators may be better alternatives (if the data support these specifications).

As quantified in the article, the inclusion of model covariates can reduce required samples, although these precision gains are tempered if the time-varying covariates are associated with treatment, which is likely to occur in practice. Thus, obtaining predictive covariates—such as prior measures of the primary study outcomes for cross-sectional cohorts—may be critical for achieving target precision levels for panel studies, especially for the CITS/ITS estimators.

Of course, sample size viability will depend on the specific study context, including the unit of treatment assignment—that often defines the study clusters—as well as data availability. We recognize that sample sizes for panel studies may be limited by the number of time periods of available data (such as from administrative records or national surveys). However, there may be some choice in the number of units selected for the study. Further, even if there is little flexibility in study sample sizes, the calculation of statistical power is still important to assess the ability of the study to detect impacts of a realistic size, and to help researchers and funders prioritize research questions that can be addressed with sufficient statistical power at reasonable cost.

# Table 1. Equation numbers of variance formulas in the text, by estimator

| Panel data estimator (relative to the average pre-period and averaged across all timing groups) | Equation number in text | |
|---|---|---|
| | Variance formula | Regression model for estimation |
| **Difference-in-differences (DID) estimator** | | |
| Pooled (average) effect in the post-period | | |
|     Cross-sectional | (14) | (7) |
|     Longitudinal | Appendix A.1 | (7), including autocorrelations for individual-level errors |
| Point-in-time effect after a certain amount of exposure to the treatment or at a specific post-period calendar time | | |
|     Cross-sectional | (15) | (7) |
|     Longitudinal | Appendix A.1 | (7), including autocorrelations for individual-level errors |
| Models with covariates | (18) | (7) with covariates |
| **Comparative interrupted time series (CITS) estimators (ITS estimators omit the comparison group terms)** | | |
| **Fully-interacted estimator** (allows intercepts and linear slopes to differ for the pre- and post-periods)[a] | | |
| Pooled (average) effect in the post-period[a] | | |
|     Cross-sectional | (24) | (19) |
|     Longitudinal | Appendix A.1 | (19), including autocorrelations for individual-level errors |
| Point-in-time effect after a certain amount of exposure to the treatment or at a specific post-period calendar time[a] | | |
|     Cross-sectional | (A.2) | (19) |
|     Longitudinal | Appendix A.1 | (19), including autocorrelations for individual-level errors |
| **Common-slopes estimator** (assumes the same linear slopes for the pre- and post-periods) | | |
| Pooled and point-in-time effects are the same | | |
|     Cross-sectional | (A.3) | (19), omitting pre-period slope terms |
|     Longitudinal | Appendix A.1 | (19), omitting pre-period slope terms and including autocorrelations for individual-level errors |

[a] The **discrete estimator** is similar to the fully-interacted estimator, except that the post-period in the model in (19) is modeled using discrete post-period indicators rather than using linear trendlines. The pooled variance estimators are the same for the discrete and fully-interacted estimators. Variances for the point-in-time discrete estimators can be obtained using (24) and footnote 5 in the main text.

**Table 2. Key notation and definitions**

| Input variable | Definition | Variable Range |
| --- | --- | --- |
| $M, M_T, M_C, r$ | Number of total, treatment, and comparison clusters, where $r = M_T/M$ is the share of treatment clusters | $M_T \geq 2$; $M_C \geq 2$ for DID / CITS designs and $M_C = 0$ for ITS design |
| $P, t$ | Total number of time periods, where time is indexed by $t$ | $P \geq 2$ for DID design; $P \geq 6$ for CITS and ITS designs |
| $N$ | Number of individuals per cluster per time period | $N \geq 1$ |
| $T_j$ | Treatment indicator that equals 1 for ever-treated clusters and 0 for comparison (never-treated) clusters indexed by $j$ | |
| $Time_t; t \in \{1, \ldots, P\}$ | Time of outcome measurement, in elapsed calendar time since a common reference point (e.g., months or quarters). These time points could be evenly or unevenly spaced. | $Time_t > 0$ |
| $K$ | Number of treatment timing groups in the treatment sample. | $K \geq 1$ |
| $S_k; k \in \{1, \ldots, K\}$ | Treatment start period for each timing group, measured using the time label, $t$, not the elapsed calendar time variable, $Time_t$. Treatment effects can be measured starting in $S_k$. | $2 \leq S_k \leq P$ for DID design; $4 \leq S_k \leq (P-2)$ for CITS / ITS designs |
| $A_k = P - S_k + 1$; $B_k = S_k - 1$ | Number of post- and pre-intervention periods | $A_k, B_k \geq 1$ for DID design; $A_k, B_k \geq 3$ for CITS / ITS designs |
| $Pre_k, Post_k$ | 1/0 indicators of pre-period ($t < S_k$) or post-period ($t \geq S_k$) | $Pre_k, Post_k = 0$ or $1$ |
| $M_{Tk}, M_{Ck}$, $p_{Tk}, p_{Ck}$ | Number of treatment and associated comparison clusters in timing group $k$, where $p_{Tk} = M_{Tk}/M_T$ and $p_{Ck} = M_{Ck}/M_C$ | $M_{Tk}, M_{Ck} > 0; \sum_k M_{Tk} = M_T$; $\sum_k M_{Ck} = M_C; \sum_k p_{\cdot k} = 1$ |
| $G_{Tj}, G_{Cj}, G_j$ | $G_{Tj} = k$ for treatment clusters in timing group $k$, $G_{Cj} = k$ for matched comparison clusters, and $G_j = k$ for both groups. | $G_{Tj}, G_{Cj}, G_j \in \{1, \ldots, K\}$ |
| $Y_{ijt}(1), Y_{ijt}(0)$ | Potential outcomes in the treated and untreated conditions for individual $i$ in cluster $j$ at time $t$ | Continuous or binary |
| $\bar{Y}(.)_{\{t=q\}}; \bar{\bar{Y}}(.)_{\{t\ldots\}}$ | Mean potential outcomes averaged over all clusters in a timing group at time $q$ ($\bar{Y}(.)$) or over a specific period ($\bar{\bar{Y}}(.)$) | Means or proportions |
| $y_{ijt}; \bar{y}_{\{t=q\}}^{G_j=k}; \bar{\bar{y}}_{\{t\ldots\}}^{G_j=k}$ | Observed outcomes and cluster-level means at a time point or period, averaged over a timing group and treatment status | Continuous or binary |
| $w_{kq}, w_k$ | Weight assigned to timing group $k$ in post-period $q$ ($w_{kq}$) or to timing group $k$ when calculating pooled effects ($w_k$) | $w_{kq} > 0; w_k = \sum_{q=S_k}^{P} w_{kq}$ |
| $\rho$ | Correlation of cluster-level outcomes over time, modeled using an AR(1) structure (or assumed constant over time) | $|\rho| < 1$ |
| $\bar{\rho}_k^{Pre}, \bar{\rho}_k^{Post}, \ldots$ | Mean correlations over specific time periods | $|\bar{\rho}| < 1$ |
| $ICC_\theta$ | Intraclass correlation coefficient measuring the percentage of total variance in the outcome that is due to variation between cluster-time cells | $0 \leq ICC_\theta \leq 1$ |
| $\psi$; $\bar{\psi}_k^{Pre}, \bar{\psi}_k^{Post}, \ldots$ | Correlation of outcomes for the same individual over time for longitudinal designs, modeled using an AR(1) structure | $|\psi| < 1$ |
| $R_{YX}^2; \mathbf{x}_{ijt}$ | Regression $R^2$ value from the inclusion of a vector of covariates, $\mathbf{x}_{ijt}$, conditional on the other model parameters | $0 \leq R_{YX}^2 < 1$; $\mathbf{x}_{ijt}$ can be continuous or binary |
| $R_{TX}^2$ | Average $R^2$ value from regressing the treatment status interaction terms on the model covariates (treatment-covariate collinearity) | $0 \leq R_{TX}^2 < 1$ |

**Figure 1. Design effects for the pooled DID estimator with staggered treatment timing and/or AR(1) errors relative to the reference design without these features**

1a: Design with staggered treatment timing and AR(1) errors relative to the reference design

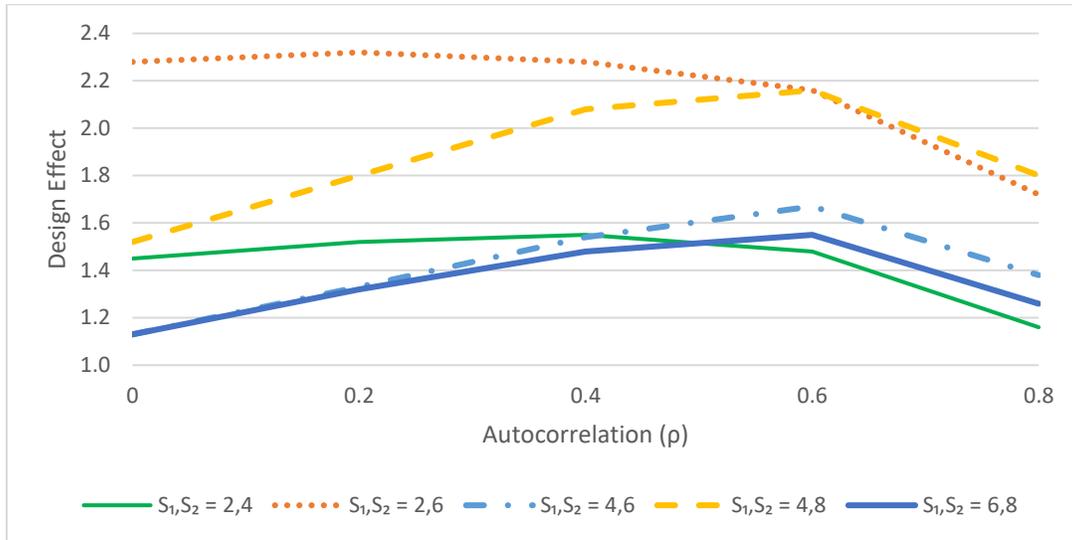

1b: Design with AR(1) errors only (no staggering) relative to the reference design

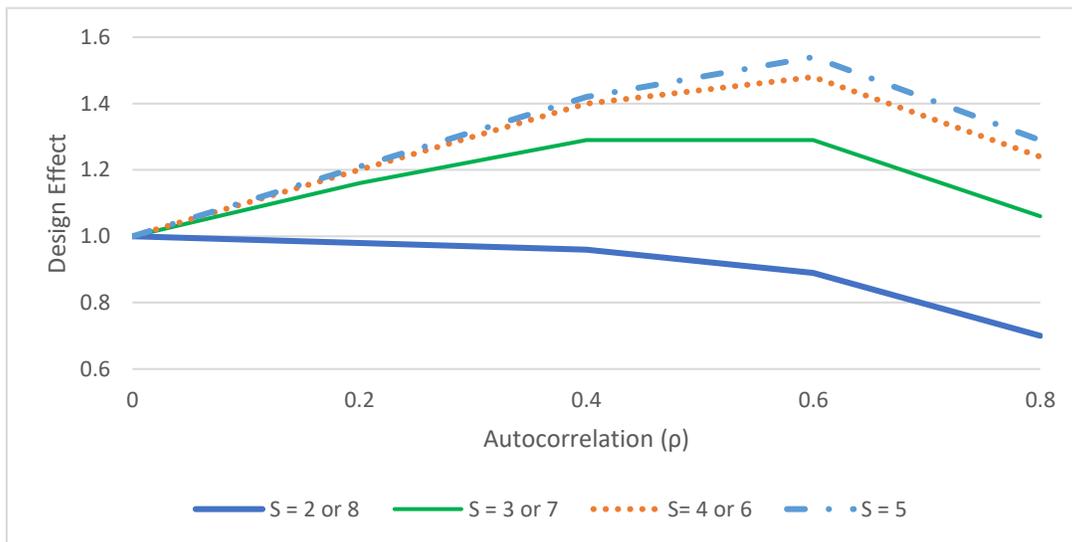

Notes. Design effects are sample inflation factors for the total number of clusters required to attain a given MDE value relative to the reference design (for a two-tailed significance test ($\alpha=0.05$) at 80 percent power), and pertain to pooled effects during the post-treatment period. Calculations assume $P = 8$ time periods, a cross-sectional design, equal time spacing of measurements, a 50-50 treatment-comparison split, $N = 100$ individuals per cluster-time cell, no model covariates, and an $ICC_\theta$ of 0.05. For the design with staggered treatment timing, the calculations assume $K = 2$ timing groups of equal size with various start times ($S_1$ and $S_2$), whereas for the reference design with $K = 1$, the treatment start date ($S$) is the average of $S_1$ and $S_2$, rounded to the nearest integer.

**Table 3. Cluster sample sizes needed to detect an *MDE* of 0.20 standard deviations**

| Number of time periods ($P$) | Treatment start times ($S_1, S_2$) | DID estimator[a] | Fully-interacted CITS estimator[a] | Fully-interacted ITS estimator[a] | Common-slopes CITS estimator[a] | Common-slopes ITS estimator[a] |
|---|---|---|---|---|---|---|
| **Pooled estimator** (cross-sectional, AR(1) errors unless indicated otherwise) | | | | | | |
| **$\rho = 0.4$** | | | | | | |
| 8 | 2,4 | 48 | NA | NA | NA | NA |
| 8 | 4,6 | 37 | 297 | 74 | 89 | 22 |
| 12 | 4,8 | 32 | 641 | 160 | 68 | 17 |
| 12 | 6,8 | 27 | 181 | 45 | 71 | 18 |
| 12 | 6,10 | 31 | 222 | 56 | 79 | 20 |
| 12 | 8,10 | 29 | 97 | 24 | 72 | 18 |
| 16 | 8,10 | 21 | 138 | 35 | 61 | 15 |
| **Constant autocorrelation ($\rho = 0.4$)** | | | | | | |
| 8 | 4,6 | 18 | 226 | 57 | 62 | 16 |
| 12 | 6,8 | 11 | 101 | 25 | 41 | 10 |
| **Longitudinal ($\rho, \psi = 0.4$)** | | | | | | |
| 12 | 6,8 | 29 | 187 | 47 | 73 | 18 |
| 12 | 6,10 | 34 | 228 | 57 | 81 | 20 |
| **Period-specific estimator** (cross-sectional, AR(1) errors) | | | | | | |
| **1 period after exposure ($\rho = 0.4$)** | | | | | | |
| 8 | 2,4 | 58 | NA | NA | NA | NA |
| 8 | 4,6 | 54 | 95 | 24 | 83 | 21 |
| 12 | 4,8 | 53 | 89 | 22 | 65 | 16 |
| 12 | 6,8 | 52 | 74 | 19 | 70 | 18 |
| 12 | 6,10 | 52 | 72 | 18 | 65 | 16 |
| 12 | 8,10 | 51 | 67 | 17 | 65 | 16 |
| 16 | 8,10 | 51 | 62 | 16 | 60 | 15 |
| **3 periods after exposure ($\rho = 0.4$)** | | | | | | |
| 8 | 2,4 | 78 | NA | NA | NA | NA |
| 8 | 4,6 | 65 | 268 | 67 | 83 | 21 |
| 12 | 4,8 | 63 | 219 | 55 | 65 | 16 |
| 12 | 6,8 | 60 | 127 | 32 | 70 | 18 |
| 12 | 6,10 | 59 | 131 | 33 | 65 | 16 |
| 12 | 8,10 | 57 | 106 | 27 | 65 | 16 |
| 16 | 8,10 | 57 | 86 | 22 | 60 | 15 |
| **5 periods after exposure ($\rho = 0.4$)** | | | | | | |
| 8 | 2,4 | 82 | NA | NA | NA | NA |
| 8 | 4,6 | 141 | 1,604 | 401 | 167 | 42 |
| 12 | 4,8 | 65 | 474 | 119 | 65 | 16 |
| 12 | 6,8 | 61 | 250 | 63 | 70 | 18 |
| 12 | 6,10 | 126 | 591 | 148 | 139 | 35 |
| 12 | 8,10 | 118 | 410 | 103 | 139 | 35 |
| 16 | 8,10 | 58 | 141 | 35 | 60 | 15 |

Notes. The figures show cluster sample sizes required to attain an MDE value of 0.20 for a two-tailed significance test ($\alpha=0.05$) at 80 percent power using the power formula in (16). Calculations assume a cross-sectional design unless otherwise noted, equal time spacing of measurements, a 50-50 treatment-comparison split, $N = 100$ individuals per cluster-time cell, no model covariates, an $ICC_\theta$ of 0.05, and $K = 2$ timing groups of equal size. NA = Not applicable because there are not enough pre-period observations for estimation.

[a] Figures pertain to the total number of treatment and comparison group clusters for the DID and CITS estimators but only to the number of treatment group clusters for the ITS estimators.

**Appendix A: Additional variance formulas**

1. **Variance formulas for the longitudinal estimators**

The variances of the DID estimators under the longitudinal AR(1) design—where the same individuals are followed over time—are similar to those for the cross-sectional estimators. The only difference is that the AR(1) structure for the individual-level error terms, $\varepsilon_{ijt}$, with autocorrelation parameter, $\psi$, must be incorporated into the analysis. To do this, note that we assume a parallel AR(1) structure for the cluster-time errors, $\theta_{jt}$, as for the individual-level errors. Thus, the adjustments to the variance formulas are parallel for both AR(1) processes, where the only difference is that there are $N$ times fewer correlations across the same individuals over time than across the same clusters over time (which affect all individuals).

Thus, we obtain the following new variance formula for the pooled longitudinal DID estimator:

$$Var(\hat{\beta}_{DID,POOL}) = \frac{1}{(\sum_{k=1}^K A_k)^2} \sum_{k=1}^K A_k^2 \left(\frac{1}{M_{Tk}} + \frac{1}{M_{Ck}}\right) \left[\sigma_\theta^2 \left(\frac{1}{A_k} + \frac{1}{B_k} + \frac{(A_k-1)}{A_k}\bar{\rho}_k^{Post}\right.\right.$$
$$\left.\left. + \frac{(B_k-1)}{B_k}\bar{\rho}_k^{Pre} - 2\bar{\rho}_k^{Pre-Post}\right) \right.$$
$$\left. + \frac{\sigma_\varepsilon^2}{N}\left(\frac{1}{A_k} + \frac{1}{B_k} + \frac{(A_k-1)}{A_k}\bar{\psi}_k^{Post} + \frac{(B_k-1)}{B_k}\bar{\psi}_k^{Pre} - 2\bar{\psi}_k^{Pre-Post}\right)\right],$$

(A.1)

where the $\bar{\psi}$ terms are defined in a parallel way as for the $\bar{\rho}$ terms provided in the main text.

The same issues for extending the variances of the cross-sectional estimators to the longitudinal context apply to all our considered DID and CITS estimators. Thus, we do not present them in this article.

## 2. Variance formulas for the point-in-time CITS estimators for the fully-interacted model

The variances of the point-in-time CITS estimators, $\hat{\beta}^e_{CITS,l}$ and $\hat{\beta}_{CITS,q}$, are provided in the following corollary to Theorem 3 presented in the main text of the article:

**Corollary to Theorem 3.** The variance formula for $\hat{\beta}^e_{CITS,l}$ that pertains to the CITS estimator after $l$ periods of treatment exposure is

$$Var(\hat{\beta}^e_{CITS,l}) = \frac{1}{(\sum_{k=1}^K I(l \leq A_k))^2} \sum_{k=1}^K I(l \leq A_k) \left[ Var(\hat{\beta}_{DID,k}) + (\frac{1}{M_{Tk}} \right.$$

$$+ \frac{1}{M_{Ck}}) \left[ \sigma^2_\theta (Term1e_k + Term2e_k + Term3e_k + Term4e_k \right.$$

$$- Term5e_k - Term6e_k - Term7e_k) + \left(\frac{\sigma^2_\varepsilon}{N}\right)(Term8e_k$$

$$\left. \left. + Term9e_k) \right] \right], \quad (A.2)$$

where

$$Term1e_k = \left(Time_{(l+S_k-1)} - \overline{Time}_{\{t \geq S_k\}}\right)^2 \left( \frac{1}{SSQT_k^{Post}} + \frac{(A_k - 1)A_k \bar{\rho}_k^{Post1}}{[SSQT_k^{Post}]^2} \right),$$

$$Term2e_k = \left(Time_{(l+S_k-1)} - \overline{Time}_{\{t < S_k\}}\right)^2 \left( \frac{1}{SSQT_k^{Pre}} + \frac{(B_k - 1)B_k \bar{\rho}_k^{Pre1}}{[SSQT_k^{Pre}]^2} \right),$$

$$Term3e_k = 2\left(Time_{(l+S_k-1)} - \overline{Time}_{\{t \geq S_k\}}\right) A_k \bar{\rho}_k^{Post2} \left( \frac{1}{SSQT_k^{Post}} \right),$$

$$Term4e_k = 2\left(Time_{(l+S_k-1)} - \overline{Time}_{\{t < S_k\}}\right) B_k \bar{\rho}_k^{Pre2} \left( \frac{1}{SSQT_k^{Pre}} \right),$$

$$Term5e_k = 2\left(Time_{(l+S_k-1)} - \overline{Time}_{\{t \geq S_k\}}\right) A_k \bar{\rho}_k^{Pre-Post2} \left( \frac{1}{SSQT_k^{Post}} \right),$$

$$Term6e_k = 2\left(Time_{(l+S_k-1)} - \overline{Time}_{\{t < S_k\}}\right) B_k \bar{\rho}_k^{Pre-Post3} \left( \frac{1}{SSQT_k^{Pre}} \right),$$

$$Term7e_k = 2\big(Time_{(l+S_k-1)} - \overline{Time}_{\{t<S_k\}}\big)\big(Time_{(l+S_k-1)}$$

$$- \overline{Time}_{\{t\geq S_k\}}\big)A_k B_k \bar{\rho}_k^{Pre-Post4}\left(\frac{1}{SSQT_k^{Post}SSQT_k^{Pre}}\right),$$

$$Term8e_k = \big(Time_{(l+S_k-1)} - \overline{Time}_{\{t\geq S_k\}}\big)^2 \left(\frac{1}{SSQT_k^{Post}}\right),$$

$$Term9e_k = \big(Time_{(l+S_k-1)} - \overline{Time}_{\{t<S_k\}}\big)^2 \left(\frac{1}{SSQT_k^{Pre}}\right),$$

$$\bar{\rho}_k^{Post1} = \frac{2}{A_k(A_k-1)}\sum_{q=S_k}^{P}\sum_{q'>q}^{P}(Time_q - \overline{Time}_{\{t\geq S_k\}})(Time_{q'} - \overline{Time}_{\{t\geq S_k\}})\rho^{(Time_{q'}-Time_q)},$$

$$\bar{\rho}_k^{Post2} = \frac{1}{A_k^2}\sum_{q=S_k}^{P}\sum_{q'=S_k}^{P}(Time_q - \overline{Time}_{\{t\geq S_k\}})\rho^{|Time_{q'}-Time_q|},$$

$$\bar{\rho}_k^{Pre1} = \frac{2}{B_k(B_k-1)}\sum_{b=1}^{B_k}\sum_{b'>b}^{B_k}(Time_b - \overline{Time}_{\{t<S_k\}})(Time_{b'} - \overline{Time}_{\{t<S_k\}})\rho^{(Time_{b'}-Time_b)},$$

$$\bar{\rho}_k^{Pre2} = \frac{1}{B_k^2}\sum_{b=1}^{B_k}\sum_{b'=1}^{B_k}(Time_b - \overline{Time}_{\{t<S_k\}})\rho^{|Time_{b'}-Time_b|},$$

$$\bar{\rho}_k^{Pre-Post2} = \frac{1}{A_k B_k}\sum_{b=1}^{B_k}\sum_{q=S_k}^{P}(Time_q - \overline{Time}_{\{t\geq S_k\}})\rho^{(Time_q-Time_b)},$$

$$\bar{\rho}_k^{Pre-Post3} = \frac{1}{A_k B_k}\sum_{b=1}^{B_k}\sum_{q=S_k}^{P}(Time_b - \overline{Time}_{\{t<S_k\}})\rho^{(Time_q-Time_b)},$$

$$\bar{\rho}_k^{Pre-Post4} = \frac{1}{A_k B_k}\sum_{b=1}^{B_k}\sum_{q=S_k}^{P}(Time_q - \overline{Time}_{\{t\geq S_k\}})(Time_b - \overline{Time}_{\{t<S_k\}})\rho^{(Time_q-Time_b)}$$

and other terms are defined in the main text. To instead calculate $Var(\hat{\beta}_{DID,q})$, we can use (A.2), replacing $I(l \leq A_k)$ with $I(q \geq S_k)$ and replacing $Time_{(l+S_k-1)}$ with $Time_q$. The ITS estimators remove the $(1/M_{Ck})$ terms.

### 3. Variance formulas for the common-slopes CITS estimator

The variance of the pooled CITS estimator with common pre- and post-period slopes can be obtained using parallel procedures as we used to calculate variances for the fully-interacted CITS estimators in the main text of the article, and is provided in the following theorem:

**Theorem 4.** The variance of the pooled, common-slopes CITS estimator $\hat{\beta}_{CITS,CS}$, that incorporates clustering, variation in intervention timing, the AR(1) error structure, and general measurement intervals is as follows:

$$Var(\hat{\beta}_{CITS,CS}) = \frac{1}{(\sum_{k=1}^{K} A_k)^2} \sum_{k=1}^{K} A_k^2 \left(\frac{1}{M_{Tk}} + \frac{1}{M_{Ck}}\right)\left(\frac{1}{A_k}\right.$$

$$\left. + \frac{1}{B_k}\right)\left[\sigma_\theta^2 (Term1CS_k + Term2CS_k - Term3CS_k + Term4CS_k) \right. \quad (A.3)$$

$$\left. + \frac{\sigma_\varepsilon^2}{N}(Term1CS_k)\right],$$

where

$$Term1CS_k = \left(\frac{SSQT^{Full}}{SSQT_k^{Pre} + SSQT_k^{Post}}\right),$$

$$Term2CS_k = \left(\frac{1}{A_k} + \frac{1}{B_k}\right)P(P-1)\left(\frac{SSQT^{Full}}{SSQT_k^{Pre} + SSQT_k^{Post}}\right)^2 \bar{\rho}_k^{Full1},$$

$$Term3CS_k = 2P(P-1)\left(\frac{SSQT^{Full}}{SSQT_k^{Pre} + SSQT_k^{Post}}\right)\left(\frac{\overline{Time}_{\{t \geq S_k\}} - \overline{Time}_{\{t < S_k\}}}{SSQT_k^{Pre} + SSQT_k^{Post}}\right)\bar{\rho}_k^{Full2},$$

$$Term4CS_k = A_k B_k (P-1)\left(\frac{\overline{Time}_{\{t \geq S_k\}} - \overline{Time}_{\{t < S_k\}}}{SSQT_k^{Pre} + SSQT_k^{Post}}\right)^2 \bar{\rho}^{Full3},$$

$$\bar{\rho}_k^{Full1} = \frac{2}{P(P-1)}\sum_{p=1}^{P}\sum_{p'>p}^{P}(Post_{kp} - \overline{Post}_{k\{t \geq 1\}})(Post_{kp'} - \overline{Post}_{k\{t \geq 1\}})\rho^{(Time_{p'} - Time_p)},$$

$$\bar{\rho}_k^{Full2} = \frac{1}{P(P-1)}\sum_{p=1}^{P}\sum_{p' \neq p}^{P}(Time_p - \overline{Time}_{\{t \geq 1\}})(Post_{kp'} - \overline{Post}_{k\{t \geq 1\}})\rho^{|Time_{p'} - Time_p|},$$

$$\bar{\rho}^{Full3} = \frac{2}{P(P-1)}\sum_{p=1}^{P}\sum_{p'>p}^{P}(Time_p - \overline{Time}_{\{t \geq 1\}})(Time_{p'} - \overline{Time}_{\{t \geq 1\}})\rho^{(Time_{p'} - Time_p)},$$

$\overline{Post}_{k\{t \geq 1\}} = (A_k/P)$ is the percentage of all periods that are post-periods for timing group $k$, and other terms are defined in the main text.

The variances for the point-in-time, common-slopes CITS estimators are similar to those for the pooled, common-slopes CITS estimator because the common-slopes model assumes a constant treatment effect over time. To obtain the variance for the point-in-time, common-slopes CITS estimator at exposure point $l$, we can use (A.3), but need to replace the weights $A_k^2$ with $I(A_k \geq l)$ and replace $(\sum_{k=1}^{K} A_k)^2$ with $(\sum_{k=1}^{K} I(A_k \geq l))^2$, and similarly for the variance of the calendar-time, common-slopes CITS estimator at post-period $q$.